\documentclass[aps,prstab,twocolumn,groupedaddress,longbibliography]{revtex4-1}
\usepackage{paralist}
\usepackage{graphicx}
\usepackage{epstopdf}
\usepackage{epsfig}
\usepackage{textcase}
\usepackage{bm}
\usepackage{url}
\usepackage{makecell}
\usepackage{multirow}

\begin{document}

\title{Design of a high power production target for the Beam Dump Facility at CERN}

\author{E. Lopez Sola}
\email[]{edmundo.lopez.sola@cern.ch}
\affiliation{CERN, 1211 Geneva 23, Switzerland}

\author{M.~Calviani}
\author{P.~Avigni}
\author{M.~Battistin}
\author{J.~Busom Descarrega}
\author{J.~Canhoto Espadanal}
\author{M.~A.~Fraser}
\author{S.~Gilardoni}
\author{B.~Goddard}
\author{D.~Grenier}
\author{R.~Jacobsson}
\author{K.~Kershaw}
\author{M.~Lamont}
\author{A.~Perillo-Marcone}
\author{M.~Pandey}
\author{B.~Riffaud}
\author{S.~Sgobba}
\author{V.~Vlachoudis}
\author{L.~Zuccalli}

\affiliation{CERN, 1211 Geneva 23, Switzerland}

\date{\today}

\begin{abstract}
The Beam Dump Facility (BDF) project is a proposed general-purpose facility at CERN, dedicated to beam dump and fixed target experiments. In its initial phase, the facility is foreseen to be exploited by the Search for Hidden Particles (SHiP) experiment. Physics requirements call for a pulsed 400 GeV/c proton beam as well as the highest possible number of protons on target (POT) each year of operation ($4.0\cdot10\textsuperscript{13}$/year), in order to search for feebly interacting particles. The target/dump assembly lies at the heart of the facility, with the aim of safely absorbing the full high intensity Super Proton Synchrotron (SPS) beam, while maximizing the production of charmed and beauty mesons. High-Z materials are required for the target/dump, in order to have the shortest possible absorber and reduce muon background for the downstream experiment. The design of the production target is one of the most challenging aspects of the facility design, due to the high energy and power density deposition that are reached during operation, and the resulting thermo-mechanical loads. The nature of the beam pulse induces very high temperature excursions between pulses (up to 100\,$^{\circ}\mathrm{C}$), leading to considerable thermally-induced stresses and long-term fatigue considerations. The high average power deposited on target (305 kW) creates a challenge for heat removal. During the BDF facility Comprehensive Design Study (CDS), launched by CERN in 2016, extensive studies have been carried out in order to define and assess the target assembly design. These studies are described in the present contribution, which details the proposed design of the BDF production target, as well as the material selection process and the optimization of the target configuration and beam dilution. One of the specific challenges and novelty of this work is the need to consider new target materials, such as a molybdenum alloy (TZM) as core absorbing material and Ta2.5W as cladding. Thermo-structural and fluid dynamics calculations have been performed to evaluate the reliability of the target and its cooling system under beam operation. In the framework of the target comprehensive design, a preliminary mechanical design of the full target assembly has also been carried out, assessing the feasibility of the whole target system.
\end{abstract}


\maketitle

\section{Introduction}
Currently in its design phase, the new Beam Dump Facility (BDF) is a multi-purpose facility aiming for high-intensity beam dump and fixed target experiments. The first objective of the facility is to explore Hidden Sector models and to search for Light Dark Matter and other feebly interacting particles with the Search for Hidden Particles (SHiP) experiment~\cite{EOI_SHiP,Anelli_SHiP,Alekhin_SHiP,Ahdida_2019}. The new facility also offers the opportunity to explore rare $\tau$ lepton decays and $\tau$ neutrino studies through fixed target flavour physics programs. In order to reach the physics objectives set forth by the SHiP experiment, a pulsed high energy beam is required coupled with the highest possible number of protons on target. The CERN accelerator complex, in particular with the Super Proton Synchrotron (SPS) beam, has been identified as the ideal machine amongst the facilities currently in operation~\cite{Anelli_SHiP}.

At the core of the facility, a dense target/dump surrounded by heavy shielding~\cite{BDFcomplex} will have a double function: (i) to absorb safely and reliably the entire energy of the 400 GeV/c Super Proton Synchrotron (SPS) beam; (ii) its design must be optimized from a physics perspective in order to maximize the production of charm and beauty hadron decays as well as photons, all of which are potential sources of very weakly coupled particles. A series of particle detectors will be situated downstream of the target complex, with the aim of searching for portal interactions with hidden sector particles, including Dark Matter candidates.

The BDF target can be considered as a target/dump assembly, since it will contain most of the particle cascade produced by the primary SPS beam interaction. This requirement leads to a very challenging target design, given the high energy and power density that will be deposited during operation, and the subsequent thermal and structural loads. The materials sought for the production target are high-A/Z materials with a short interaction length, aiming at maximizing the re-absorption of pions and kaons produced in the intra-nuclear cascade process, which are particles considered as background for the SHiP experiment. As an additional requirement for the material selection, the target materials shall exhibit suitable physical and mechanical properties to withstand the beam-induced stresses on the target over its entire lifetime under severe conditions (irradiation, high temperatures and high-cycle fatigue).

The 400~GeV/c proton beam pulse from the SPS is foreseen to impact the target with a pulse duration of one second, delivering an average power on target of 2.56 MW, followed by a cooling of 6.2 seconds. This configuration leads to a beam average power of 355 kW, of which roughly 305 kW will be dissipated inside the target assembly, while the rest will be lost in the surrounding shielding of the BDF target complex. A detailed list of the BDF operation beam parameters is specified in Table~\ref{tab:beamparameters}. Due to the pulsed nature of the SPS machine (0.16 Hz), the instantaneous power per pulse is unprecedented for similar installations based on solid high-Z targets. As a comparison, the ISIS neutron sources~\cite{THOMASON201961} are operating at frequencies of 50 Hz (TS-1~\cite{Bungau:2014nda}) or 10 Hz (TS-2~\cite{DEY201863}), while the LANSCE Lujan Center production target is operating at 20 Hz~\cite{NOWICKI2017374}. Taking into account similar power deposited on target, the beam configuration leads to higher transients with respect to the cited facilities, considered as continuous or quasi-continuous sources.

Liquid targets have been excluded from this study due to safety considerations and very limited long-term operational experience. 

The target is expected to survive for 5 years of operation at $4.0\cdot10\textsuperscript{19}$ protons on target per year, corresponding to a total of $2.0\cdot10\textsuperscript{20}$ protons over the lifetime of the target.

\begin{table}[hbtp]
\caption{\label{tab:beamparameters}%
Baseline beam parameters of the BDF target operation.}
\begin{ruledtabular}
\begin{tabular}{lc}
\multicolumn{2}{c}{Baseline characteristics}\\
\hline
Proton momentum {[}GeV/c{]}              & 400                        \\
Beam intensity {[}p$^+$/cycle{]}            & $4.0\cdot10\textsuperscript{13}$ \\
Cycle length {[}s{]}                     & 7.2                        \\
Spill duration {[}s{]}                   & 1.0                        \\
Beam dilution pattern [-] & Circular \\
Beam sweep frequency [turns/s] & 4\\
Dilution circle radius [mm] & 50 \\
Beam sigma (H,V) [mm] & (8,8) \\
Average beam power  {[}kW{]}    & 356                       \\
Average beam power deposited on target {[}kW{]}    & 305                        \\
Average beam power during spill {[}MW{]} & 2.3                       \\ 
\end{tabular}
\end{ruledtabular}
\end{table}

\section{BDF Target design}
\label{sec:design}

\subsection{Production target material selection}
\label{Sec:TGT:Design:Material}
Targets and dumps were built in the past for similar physics objectives, such as those for the DONUT experiment at Fermilab~\cite{PhysRevD.78.052002} or for the CHARM experiments at CERN~\cite{DORENBOSCH1986473,doi:10.1111/j.1749-6632.1988.tb51542.x}. In those experiments, however, the target/dumps were designed to receive much lower beam pulse intensities (around few 10$^{12}$) and average beam power (maximum of around 20 kW average beam power), than those foreseen for BDF. These conditions led to lower instantaneous temperature rises as well as reduced steady-state temperatures and resulting stresses.
In addition, the total number of protons on target for both DONUT and CHARM experiments, in the order of 10$^{17}$ POT, was almost 3 orders of magnitude lower than the foreseen value for the BDF target. Therefore, the BDF target/dump assembly requires a completely novel approach, with different configurations and material technologies to safely operate with the new challenges.

\begin{figure}
\resizebox{0.45\textwidth}{!}{
\includegraphics{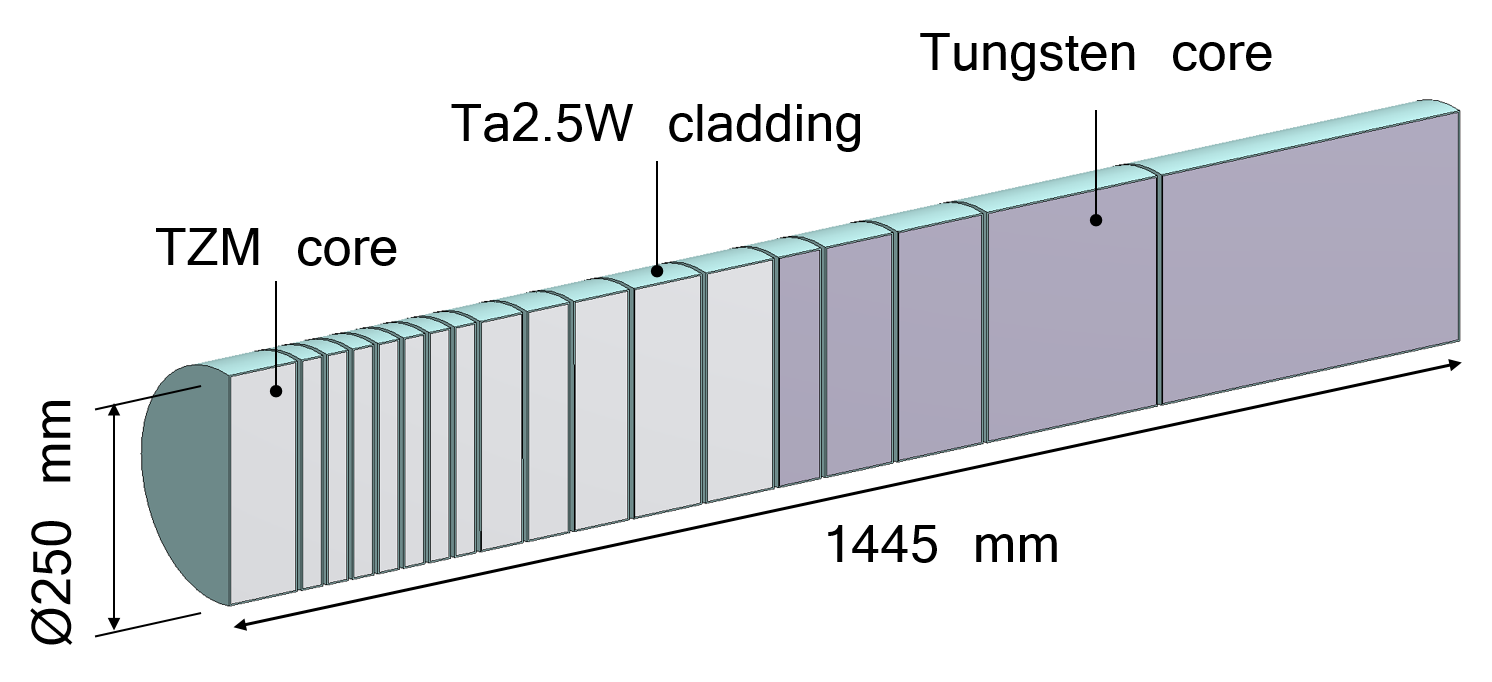}}%
\caption{\label{fig:TGT:BDFtarget} Layout of the Beam Dump Facility target core. The target blocks are made of TZM or pure tungsten cylinders clad with a tantalum alloy, Ta2.5W.}
\end{figure}

The proposed target is a hybrid assembly, consisting of several collinear cylinders of TZM ((0.08\%) titanium - (0.05\%) zirconium - molybdenum alloy) and pure tungsten (W), clad with a W-containing Ta-alloy, Ta2.5W ((2.5\%) tungsten - tantalum alloy), see Figure \ref{fig:TGT:BDFtarget}. As a result, the target has a total number of nuclear interaction lengths of $\sim$12$\lambda$. A target core fully made of pure W has been excluded due to the excessive stresses that would be present in the first part of the target, where the energy deposition is the highest. 

\begin{itemize}
    \item TZM is chosen for the first part of the target core, that will absorb most of the beam power deposited on the target. This material has a density high enough to fulfill the experiment requirements, but leading to an energy deposition lower than for a denser material such as pure tungsten. This \mbox{Mo-based} alloy is chosen because it features higher strength, better creep resistance and higher recrystallization temperature compared to pure molybdenum~\cite{TantalumW2}.
    \item  For the second part of the target, that will receive a lower amount of power deposition from the primary beam, pure W is selected, since it fulfills the physics requirements (high Z-number and short interaction length), and has proven good performance under irradiation~\cite{tungstenprops}.
\end{itemize}

The target needs to be actively cooled: the authors have considered different options including water and helium gas (as implemented at ESS~\cite{Garoby_2017}). Nevertheless, given the high energy deposited and the high temperatures reached during operation, priority was given to the design and implementation of a water-cooled target. CERN has a large experience in cooling beam intercepting devices with water, with appropriate means in place to deal with tritium production in cooling loops. The cooling system design is based on the circulation of a high velocity water stream through 5 mm gaps foreseen between the different target blocks. More details about the cooling system design are given in Section~\ref{Sec:TGT:CoolingCFD}. 

The high velocity water flow in contact with the pure W and TZM blocks could induce undesired corrosion-erosion effects. Therefore, all the target core blocks will be clad via diffusion bonding achieved by means of the Hot Isostatic Pressing (HIPing) method with Ta2.5W~\cite{HIP1,HIP2,HIP_Busom}. This material is selected as cladding material due to its high corrosion resistance and its convenience as high-Z material with short interaction length. Ta2.5W has enough strength and ductility to withstand the HIP conditions necessary for diffusion bonding (temperature and pressure), and is soluble with molybdenum and tungsten, so there is no risk of forming any intermetallic alloy during the HIP diffusion bonding.

In the preliminary target design phase, pure tantalum was considered as cladding material for the target core blocks, given the vast experience with tantalum-clad targets in other operating facilities such as the ISIS target stations at the Rutherford Appleton Laboratory (RAL)~\cite{ISISclad}. However, due to the instantaneous power deposited on target during the spill, the structural calculations performed (detailed in Section~\ref{Sec:TGT:Simus}) have shown that the maximum stresses reached in a pure tantalum cladding may be critical for the target operation, limiting its lifetime considerably. For that reason, Ta2.5W has been considered as alternative cladding material, with the advantage of a higher strength at high temperatures and a similar corrosion-erosion resistance~\cite{TantalumW2,TantalumW}.

A detailed R\&D study has been carried out in order to test the bonding quality of Ta2.5W with TZM and pure tungsten after the HIP process. The interface mechanical tests performed have proven that the intermetallic bonding strength of TZM or W with Ta2.5W is comparable to the one with pure tantalum, validating the selection of Ta2.5W as target cladding material~\cite{HIP_Busom}.

To further consolidate the choice of Ta2.5W for the BDF target blocks cladding, a prototype of the BDF target has been tested under beam in the North Area of CERN~\cite{sola2019beam}. The target prototype consists in a scale replica of the BDF target, with identical length and reduced diameter. Pure tantalum and Ta2.5W have been used as cladding materials for the target prototype, in order to compare the performance of both materials under beam irradiation. A Post Irradiation Examination (PIE) campaign is foreseen during 2020 on several blocks of the target prototype, to characterize the mechanical bonding of the cladding and core materials after irradiation.

\subsection{Optimization of the target design}
\label{sec:design:optimization}

The BDF target core is made of 18 collinear cylinders with a diameter of 250~mm and variable thicknesses, from 25~mm to 80~mm for the TZM blocks and from 50~mm to 350~mm for the pure tungsten blocks, giving a total effective target core length of around 1.3~m. The target cylinders length has been iteratively adjusted to reduce the level of temperatures and stresses reached in the different materials. Table~\ref{tab:TGT:targetblocks} summarizes the target core materials and the blocks longitudinal thickness. This configuration is a result of a series of iterations balancing the physics requirements of the SHiP experiment~\cite{EOI_SHiP,Anelli_SHiP,Alekhin_SHiP,Ahdida_2019} and the survivability of the target/dump assembly.

\begin{table}[hbtp]
\caption{\label{tab:TGT:targetblocks} Summary of the BDF final target cylinders longitudinal thickness and materials}
\begin{ruledtabular}
\begin{tabular}{cccc}
Block number & Core material & Length (mm) & Weight (kg) \\
\hline
1            & TZM                       & 80      &40    \\
2            & TZM                       & 25       &12.5   \\
3            & TZM                      & 25      &12.5   \\
4            & TZM                       & 25      &12.5    \\
5            & TZM                      & 25       &12.5    \\
6            & TZM                      & 25       &12.5    \\
7            & TZM                     & 25      &12.5    \\
8            & TZM                      & 25     &12.5     \\
9            & TZM                       & 50     &25     \\
10           & TZM                     & 50         &25 \\
11           & TZM                     & 65          &33\\
12           & TZM                & 80      &40    \\
13           & TZM                   & 80      &40    \\
14           & W                     & 50       &47   \\
15           & W                        & 80       &76   \\
16           & W                        & 100       &95  \\
17           & W                       & 200    &190     \\
18           & W                      & 350     &330    \\
\end{tabular}
\end{ruledtabular}
\end{table}

\begin{figure}
\resizebox{0.45\textwidth}{!}{
\includegraphics{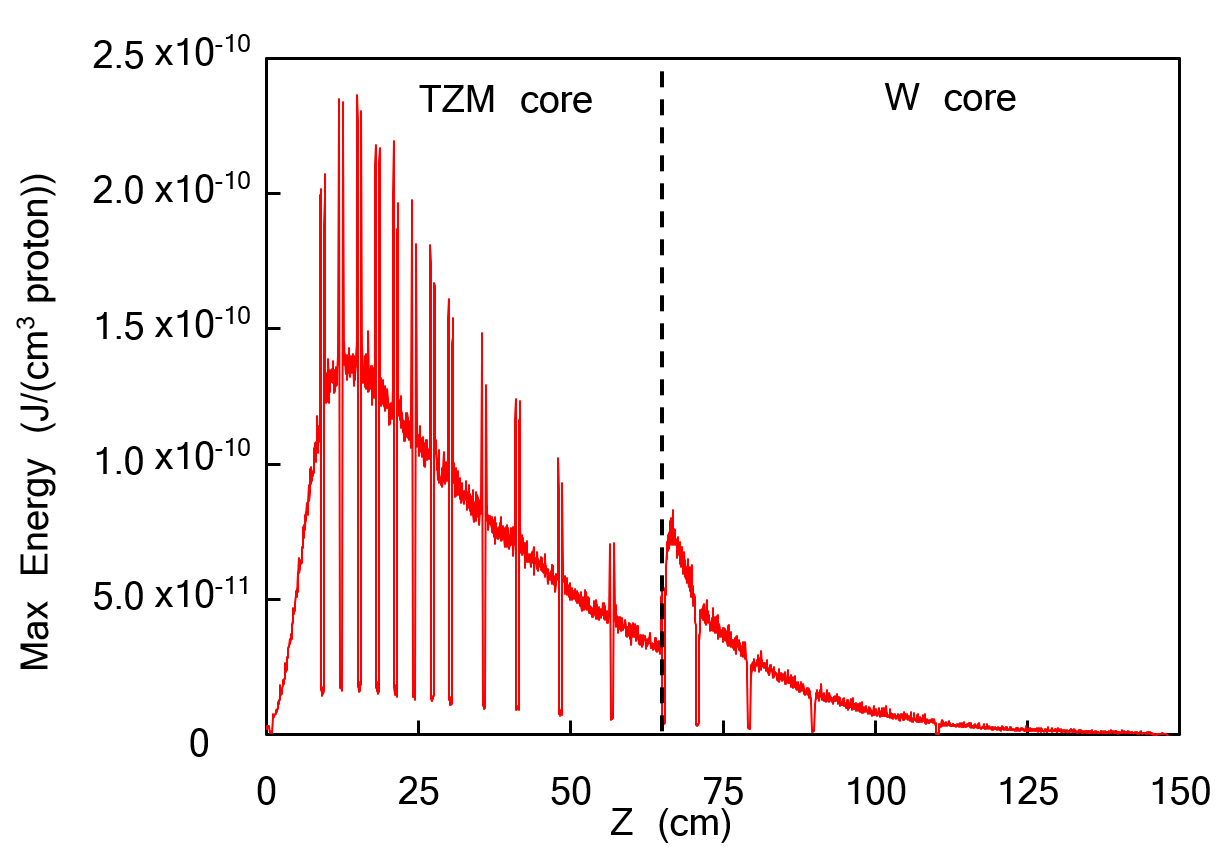}}
\caption{\label{fig:TGT:FLUKAenergy} Maximum deposited energy density per proton in the BDF target blocks along the longitudinal axis, obtained via FLUKA Monte Carlo simulations~\cite{FLUKA_Code}.}
\end{figure}

Figure~\ref{fig:TGT:FLUKAenergy} illustrates the maximum deposited energy density in the longitudinal direction normalized per proton impact on target, calculated via FLUKA Monte Carlo simulations~\cite{FLUKA_Code}. Given that Ta2.5W has a higher density than the one of TZM, the highest energy density values are reached in the cladding of the target blocks. The lower peaks correspond to the water gaps between the blocks. 

The optimization of the thickness of the target blocks is based on the distribution of maximum energy deposition per block as well as on the maximum physics reach. The maximum energy values for the TZM part (first half of the target) are found in blocks 2 to 8, therefore these blocks have been segmented to have the shortest length (25~mm). The reduced thickness of these blocks allows for a more effective heat dissipation by the water flowing through the 5~mm gaps between the target cylinders, hence reducing the maximum temperatures. 

The thickness of the following blocks is gradually increased as the total deposited energy density per block is decreasing. An identical approach is used for the second part of the target (pure tungsten core): the first block is the most loaded one in terms of heat deposition, hence the thickness of this block is set to 50 mm. The length of the following tungsten core blocks is then increased progressively as the deposited energy density decreases.

As will be discussed in Section~\ref{Sec:TGT:Simus}, the most critical cladding temperatures and stresses will be reached in the upstream thin blocks (i.e. blocks 3 to 6), where the interaction with the primary beam leads to the highest values of deposited energy density. A further reduction of the target operating temperature would be possible through an additional segmentation of the low-thickness target blocks. However, the insertion of supplementary water gaps in the target design is not desirable from the physics perspective, since the presence of water gaps in the longitudinal direction reduces the effective stopping power of pions and kaons, and consequently increases the neutrino background from pion and kaon decays.

\subsection{Dilution system requirements for target}
\label{Sec:TGT:Design:dilution}

The high energy deposited on target requires dilution by sweeping the beam over the target during the spill using upstream magnets \cite{EDMS-ExtractionSHIP}, since the impact of a non-diluted beam would lead to premature failure of the target. The beam dilution pattern has been optimized taking into account the mechanical performance of the target and the different restrictions imposed by the magnets composing the dilution system in the transfer line. As a result, the SPS primary beam is foreseen to be swept following a circular pattern, with 4 turns over a 50 mm radius circle for each one second pulse. The maturity of the current configuration is the result of an iterative process. 

Several beam dilution failure scenarios have been studied in detail, and the implementation of methods for failure detection and prevention has been analysed~\cite{YellowBook}.

The beam size has also been maximized taking into account the limitations imposed by the aperture of the upstream transfer line magnets. Larger beam spot sizes lead to lower temperatures and stresses on target, since the energy deposition is more distributed in the material volume. The compromise between maximizing the spot size of the round beam and the aperture restrictions of the transfer line magnets concluded with the selection of a beam size on target of 8~mm (1$\sigma$). 

\section{Target thermo-mechanical calculations}
\label{Sec:TGT:Simus}

\subsection{Thermal calculations}
\label{Sec:TGT:Simus:thermal}

The energy deposited by the primary proton beam on the target is evaluated with help of FLUKA~\cite{FLUKA_Code} Monte Carlo simulations. The energy induced by beam-matter interactions is imported into a Finite Element Analysis (FEA) software (ANSYS Mechanical~\textsuperscript{\copyright}) in order to evaluate the target performance during operation.

Forced convection has been applied on the surfaces of the target blocks as a boundary condition for the FEM thermal simulations, with a film coefficient value of 20000 W/(m\textsuperscript{2 }K) and a water temperature of $30\,^{\circ}\mathrm{C}$. These values have been estimated through analytical calculations and are consistent with the average heat transfer coefficient (HTC) calculated via Computational Fluid Dynamics (CFD) simulations, as shown in Section~\ref{Sec:TGT:CoolingCFD}.

\begin{figure*}
\resizebox{0.8\textwidth}{!}{
\includegraphics{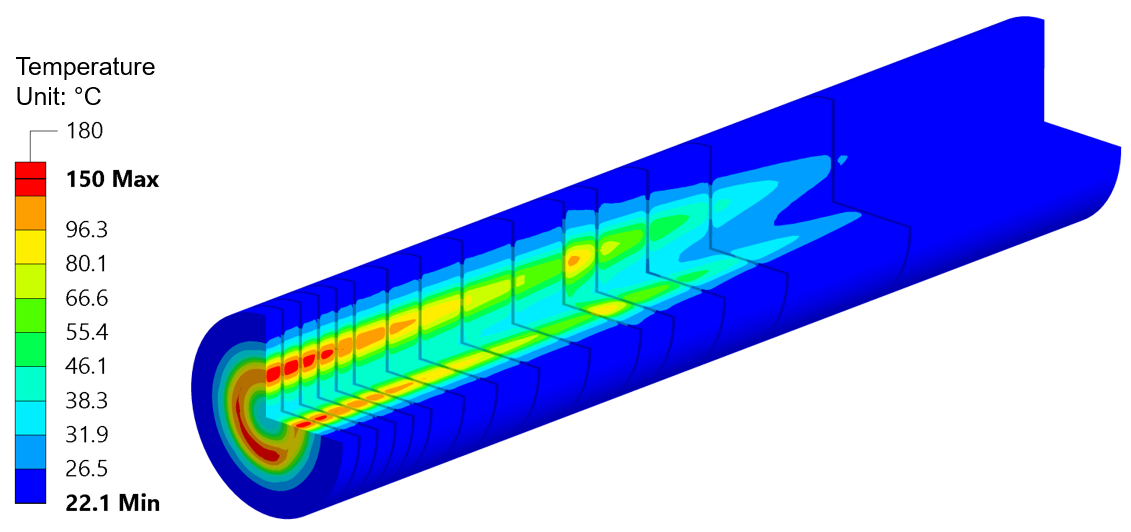}}
\caption{\label{fig:TGT:dilution_temp} Temperature distribution in the BDF target at the end of the one second long spill. The effect of the BDF beam dilution system can be clearly appreciated: four beam sweep turns in one second over a 50 mm radius circular pattern.}
\end{figure*}

One of the most challenging aspects of the thermal calculations performed is the implementation of the proton beam dilution into the FEM software. An ANSYS\textsuperscript{\copyright} Parametric Design Language (APDL) code has been developed to simulate the beam sweep trajectory following a circular pattern identical to the dilution sweep design. As a result, a time-dependent temperature distribution in the target blocks is obtained (Figure~\ref{fig:TGT:dilution_temp}). Table~\ref{tab:TGT:maxtemps} summarizes the maximum temperatures reached in the different target materials for the most critical blocks in terms of thermal loads for each of the employed target materials. 

\begin{table}[hbtp]
\begin{ruledtabular}
\caption{\label{tab:TGT:maxtemps} Maximum temperatures expected in the different BDF target materials for the most critical blocks.}
\begin{tabular}{lcc}
Material & Block number & {Maximum temperature} \\
\hline
Ta2.5W   & 4            & $160\,^{\circ}\mathrm{C}$              \\
TZM      & 9            & $180\,^{\circ}\mathrm{C}$              \\
W        & 14           & $150\,^{\circ}\mathrm{C}$        \\
\end{tabular}
\end{ruledtabular}
\end{table}

\begin{figure}
\resizebox{0.45\textwidth}{!}{
\includegraphics{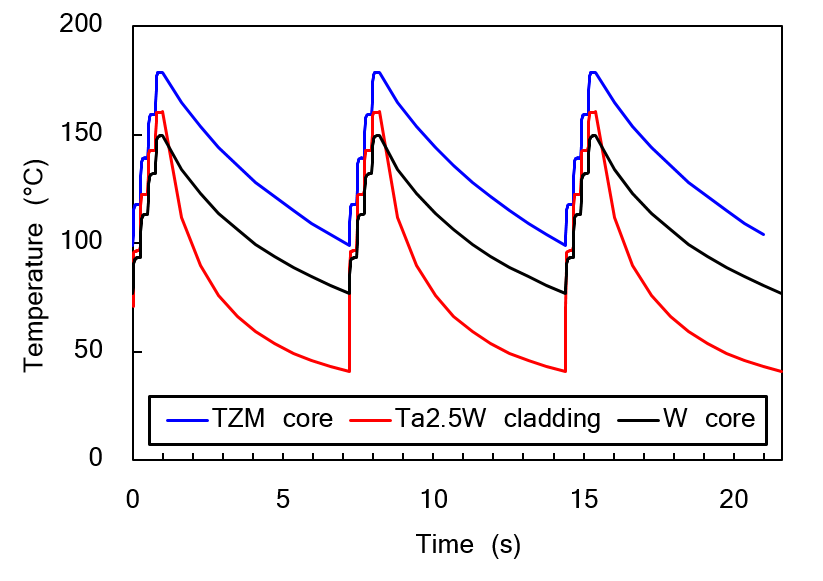}}
\caption{\label{fig:TGT:thermalplot} Maximum temperature evolution during three beam pulses after long-time operation for the BDF target materials, starting from a steady state condition. The results are shown for the most loaded target blocks.}
\end{figure}

Figure~\ref{fig:TGT:thermalplot} shows the maximum temperature evolution during three pulses for TZM, W and Ta2.5W, starting from a steady state condition. The effect of the beam dilution can be observed as the temperature increases in four steps during the beam pulse.

The temperature reached in the core and cladding materials is around 0.1 T$_{\text{m}}$ (melting temperature in K); these materials do not present any allotropic transformation in this range of temperatures. As a consequence, the evolution of the physical and mechanical properties with temperature is expected to evolve gradually, without abrupt changes. However, there are several limitations associated with the high temperatures expected during operation. 

First, the degradation of the material properties at high temperatures, specially for Ta2.5W, leads to a reduction of the strength of the material (see Section~\ref{Sec:TGT:Simus:struct}). Furthermore, it is undesirable to reach temperatures above the boiling point of water on the target surface, which could lead to localized boiling of the cooling water, inducing a severe degradation of the heat dissipation from the blocks. This issue will be reported in more detail in Section~\ref{Sec:TGT:CoolingCFD}. Finally, the thermal loads applied to the target are responsible for high levels of stresses, in particular for the cladding material where the increase of temperature after each proton beam impact can reach up to $120\,^{\circ}\mathrm{C}$, as shown in Figure~\ref{fig:TGT:thermalplot}.

\subsection{Structural calculations}
\label{Sec:TGT:Simus:struct}

\subsubsection{Target materials properties review}

For an accurate simulation of the materials behaviour during the target operation, it is important to take into account the changes in the material properties as a function of temperature. The strength of the target materials is a crucial parameter when estimating the target lifetime, and thus, when designing the target assembly.

As a general trend, for the three materials, the yield and tensile strength decrease with temperature~\cite{TZM_Filacchioni,Tungsten_Schmidt,TaW_HCStarck}. Ta2.5W shows much lower strength values than TZM and W over the entire range of temperatures. It must be noted that the differences in mechanical properties between products of the same material can be significant and they highly depend on the geometry and manufacturing process applied. 

The most relevant values of yield and tensile strength that have been measured or found in literature are presented below:

\begin{itemize}
    \item The yield and tensile strength of Ta2.5W were measured at room temperature for specimens obtained from Ta2.5W disks treated with a HIP cycle identical to the one used to produce the BDF target blocks. The measured yield strength was 270 MPa, and the tensile strength 360 MPa~\cite{TaW_yieldFH}. For comparison purposes, pure tantalum specimens with identical geometry have also been tested after the HIP process, showing much lower yield strength (170 MPa) and tensile strength (220 MPa). This low strength exhibited by pure tantalum would probably compromise the target lifetime as mentioned in Section~\ref{Sec:TGT:Design:Material}.
    
    Yield strength values of Ta2.5W sheets at the target operational temperatures can be also obtained from Ref.~\cite{TaW_HCStarck}, which reports values at room temperature which are similar to the ones measured for Ta2.5W disks post-HIP process. Values at high temperature are also reported: the expected yield strength of Ta2.5W at 200$^\circ$C is 190 MPa.
    
    \item Regarding {TZM}, samples obtained from a 200 mm diameter, 100 mm long rod were tested at 20 and $700\,^{\circ}\mathrm{C}$~\cite{TZM_yieldPlansee}. At room temperature, the measured yield strength and mean tensile strength were 480 and 525 MPa respectively. At $700\,^{\circ}\mathrm{C}$, the measured yield strength is 290 MPa. These values are significantly lower than the ones usually found in literature, as in Ref.~\cite{TZM_Filacchioni}. This is probably due to the bigger grain size obtained in the production of rods with such a large diameter and length, which is also the case of the TZM rods produced for the BDF target. Considering the reduction of material strength with temperature reported in Ref.~\cite{TZM_Filacchioni}, the estimated yield strength at $200\,^{\circ}\mathrm{C}$ is assumed to be around 430 MPa.
    
    \item {Pure tungsten} presents brittle behaviour at room temperature, and for common commercial tungsten products the ductile-to-brittle-transition-temperature (DBTT) is around $300\,^{\circ}\mathrm{C}$ in most cases~\cite{W_DBTT_Zhang,W_DBTT_2}. It is therefore foreseen that the BDF target tungsten will operate in the brittle regime, as the maximum temperatures are expected to be around $150\,^{\circ}\mathrm{C}$. Tensile testing on tungsten specimens produced via sintering and HIP has been carried out in Ref.~\cite{Wfatigue}, measuring an Ultimate Tensile Strength (UTS) at room temperature of 570 MPa. This value is relevant to the BDF target material properties study, given that the large diameter and length of the BDF tungsten blocks will most probably constrain the choice of the tungsten cylinders' production procedure to the sintering and HIP method instead of forging or rolling (see Section~\ref{Sec:TGT:MechDesign}).
    
    A reduction of the tensile strength of tungsten at $150\,^{\circ}\mathrm{C}$ to 60\% of the UTS at room temperature has been reported in~\cite{Tungsten_Schmidt}. Taking into account this reduction of strength, the estimated UTS at $150^{\circ}\mathrm{C}$ is assumed to be around 330 MPa.
\end{itemize}

From a radiation damage standpoint, the target materials are expected to receive up to 0.5 DPA over the expected lifetime of the facility. Due to the beam dilution on target, a large volume of the core will be affected by radiation damage, e.g. more than 4000 cm$^3$ are expected to receive above 0.05 DPA per year. Given that a bulk material damage (and not only a superficial or limited volume damage) is foreseen, the effect of irradiation on the target materials has been identified as a potential issue for the target operation. 

The annual gas production in the target materials has been estimated with FLUKA calculations. In a year of operation, a maximum of about 40 atomic parts per million (appm) of hydrogen and 15 appm of helium are expected in the TZM and W cores. It is foreseen that the Ta2.5W cladding will receive slightly higher DPA and H/He production -- around 50 appm of hydrogen and 30 appm of helium in one year -- given its location in the blocks outside surface and its higher density.

It is expected that TZM and tungsten will undergo radiation hardening with a pronounced brittle behavior after a few years of operation~\cite{W_radiation, TZM_radiation}. The increase of strength is regarded as beneficial, but the materials embrittlement could be harmful for the target lifetime, due to easier crack propagation and higher sensitivity to defects. A high safety factor is required for these materials in terms of static stress and fatigue life in order to increase operational robustness. Ta2.5W has been reported to maintain its ductility and increase slightly its strength after irradiation, especially if high purity material is used~\cite{Ta_radiation_BYUN,Ta_radiation_CHEN}.

Further studies are ongoing to evaluate the effect of irradiation on the thermo-physical properties of the BDF target materials. As an example, the degradation of the thermal conductivity of tungsten after proton irradiation has been reported in Ref.~\cite{TC_tungsten}, and it shall be studied if the Coefficient of Thermal Expansion (CTE) of the target materials will also be affected in a detrimental manner by radiation damage. An irradiation program has been launched in order to assess this within the framework of the RaDIATE Collaboration, focusing on TZM and Ta2.5W~\cite{Hurh:IPAC2017-WEOCB3,Ammigan:IPAC2017-WEPVA138}.

\subsubsection{Finite Element Model (FEM) simulation results}

The stresses induced by the high temperatures reached in the target materials have been estimated by means of FEM calculations. The simulations performed have shown that the level of stresses is substantially reduced if the target blocks are allowed to expand freely after the temperature rise generated by the beam impact. Therefore, in the analysis the target cylinders are considered to be resting on the support (not fixed or constrained). The required gap between the different blocks necessary for the water passage is ensured by 5 mm spacers added to the support; these maintain the blocks in position while respecting the minimum tolerance required to avoid any mechanical constraint during the blocks thermal expansion. A detailed description of the target mechanical assembly will be given in Section~\ref{Sec:TGT:MechDesign}. 

The temperature distribution for each of the blocks has been imported as a thermal load for the structural analysis. A transient structural analysis has been performed to evaluate the stress evolution over time for a given temperature distribution at each time step. The slow application of thermal loads (due to the pulse duration of one second) allows inertia effects to be neglected; for that reason the structural analysis can be regarded as quasi-static.

Ta2.5W and TZM present ductile behaviour at the target's operational temperatures, and if the yield points of the materials are reached, the blocks will start deforming in the plastic regime. However, the cyclic plastic deformation of the core or the cladding during a long period could lead to premature fracture of the material and/or to detachment of the cladding with respect to the base refractory metal, reducing or blocking the heat dissipation through the cladding material. Therefore, plastic deformation is undesirable, and the von Mises yield criterion has been used to evaluate the safety margin of the stresses reached in the Ta2.5W cladding and TZM core with respect to the temperature-dependent yield strength of these materials. 

After a few years of operation, embrittlement of the TZM core is expected due to radiation damage; no plastic deformation is foreseen at that point for TZM. Since the TZM strength will also be increased, the comparison of the von Mises equivalent stress with the yield strength under un-irradiated conditions is assumed to be a conservative approach.

For pure tungsten, which is considered as a brittle material at the target operational temperatures, the Christensen criterion~\cite{Christensen} is considered the most suited failure criterion. However, due to the low availability of compressive strength data for tungsten under the target operational conditions, the maximum normal stress criterion has been used to assess if the maximum stresses reached in the tungsten core are within the safety limits of the material.

Table~\ref{tab:TGT:maxstress} summarizes the maximum stresses calculated for each material, as well as the safety margin with respect to the yield strength or ultimate tensile strength of the material at the operational temperatures.

\begin{table}[hbtp]
\caption{\label{tab:TGT:maxstress} Maximum stresses reached in the BDF target materials and safety factor with respect to the material limits reported in literature~\cite{TaW_yieldFH,TZM_yieldPlansee,Wfatigue}.}
\begin{ruledtabular}
{\renewcommand{\arraystretch}{1.2}%
\begin{tabular}{cccc}
{Material} & {\begin{tabular}[c]{@{}c@{}}Max. Von Mises\\eq. stress $\left[\text{MPa}\right]$\end{tabular}}& {\begin{tabular}[c]{@{}c@{}}Yield strength\\ at $200\,^{\circ}\mathrm{C}$ $\left[\text{MPa}\right]$\end{tabular}} & {\begin{tabular}[c]{@{}c@{}}Safety \\ factor\end{tabular}} \\
\hline
{TZM}& 128 & 430 & 3.5 \\
{Ta2.5W} & 95 &  190 & 2 \\
\hline\\
\hline
{Material} & {\begin{tabular}[c]{@{}c@{}}Max. principal\\stress $\left[\text{MPa}\right]$\end{tabular}}& {\begin{tabular}[c]{@{}c@{}}UTS\\at $150\,^{\circ}\mathrm{C}$ $\left[\text{MPa}\right]$\end{tabular}} & {\begin{tabular}[c]{@{}c@{}}Safety \\ factor\end{tabular}} \\
\hline
{W}& 80 & 330 & 4 \\
\end{tabular}}
\end{ruledtabular}
\end{table}

\begin{figure*}[htbp]
\resizebox{0.95\textwidth}{!}{
\includegraphics{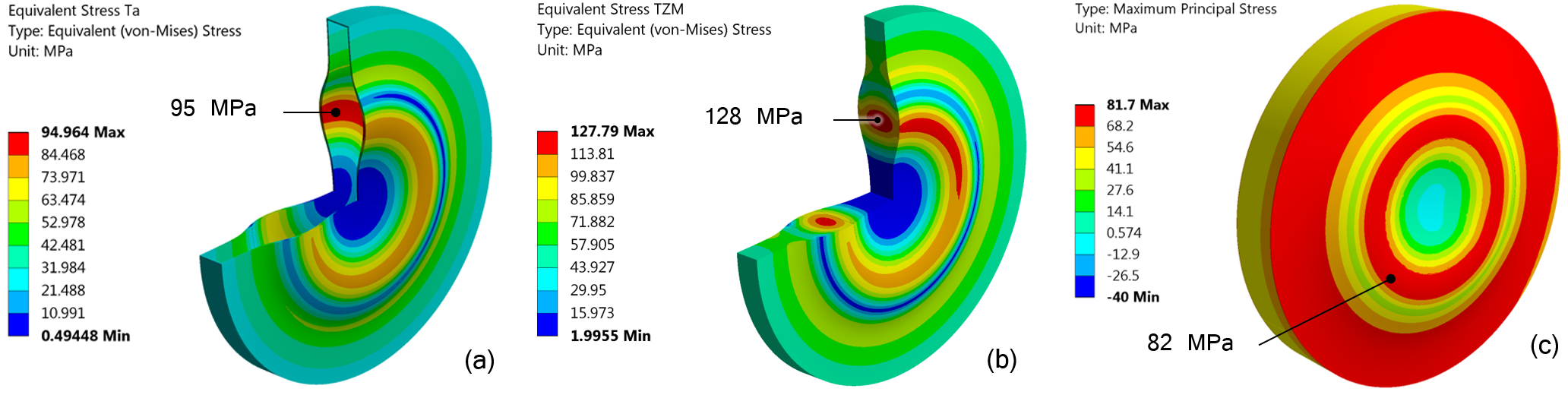}}
\caption{\label{fig:TGT:vmstresses} Von Mises equivalent stress or maximum principal stress distribution after one beam pulse in the most loaded blocks for each target material. (a) Ta2.5W cladding of block \#4, maximum equivalent stress (95 MPa) reached in the beam impact region at the interface with the block core. (b) TZM core of block \#4, maximum equivalent stress (130 MPa) found in the core centre (mainly due to high compressive stresses) and in the interface with the tantalum cladding. (c) W core of block \#14, highest value of maximum principal stress reached in the upstream face of the block, following the beam dilution path.}
\end{figure*}

\begin{figure}[ht!]
\resizebox{0.45\textwidth}{!}{
\includegraphics{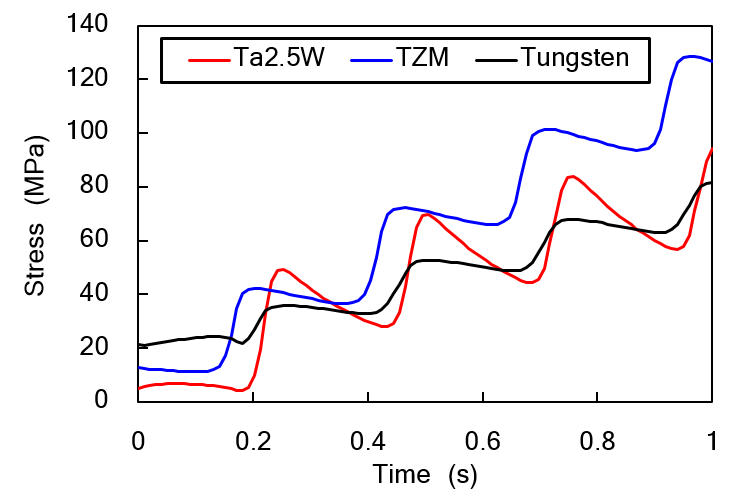}}
\caption{\label{fig:TGT:stress_evolution_sweep} Evolution during the one second SPS beam pulse of the von Mises equivalent stress (TZM and Ta2.5W), and maximum principal stress (pure tungsten). The results have been plotted for the locations of maximum stress shown in Figure~\ref{fig:TGT:vmstresses}.}
\end{figure}

The stress distribution in the most loaded blocks is shown in Figure~\ref{fig:TGT:vmstresses}, where the influence of the beam dilution system can be clearly appreciated. The beam sweep following four circular turns is also noticeable in Figure~\ref{fig:TGT:stress_evolution_sweep}: the stress value in a particular location increases as the beam approaches the analyzed point and decreases as the beam moves away.

\begin{figure}[ht!]
\resizebox{0.45\textwidth}{!}{
\includegraphics{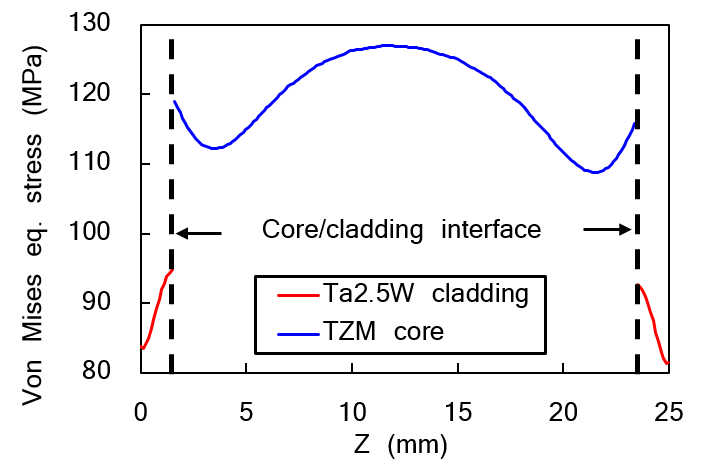}}
\caption{\label{fig:TGT:vmstressZ} Stress distribution in the longitudinal (Z) axis at the beam impact position (Y = 50 mm) for block \#4. Stress calculated at the end of the one second beam pulse on target.}
\end{figure}

As discussed in Section~\ref{sec:design:optimization}, the largest values of stress are found in the upstream thin blocks (3 to 6). The stress distribution inside one of the most critical target blocks is displayed in Figure~\ref{fig:TGT:vmstressZ}, showing an abrupt variation of the von Mises equivalent stress in the interface between the target cladding and the core. For all the target materials, the maximum stresses are well within the material limits. The target Ta2.5W cladding, with a lower safety factor with respect to the design criteria, is considered to be one of the most critical parts of the target.

\subsection{Fatigue analysis}
\label{sec:simus:struct:fatigue}

\subsubsection{Fatigue literature review}

The SPS primary beam is foreseen to impact the BDF target for one second every 7.2 seconds, and the total number of cycles expected during the target lifetime is of the order of 10\textsuperscript{7}. This means that the target will operate under cyclic structural loads, and it is necessary to evaluate the fatigue life of the target materials under high-cycle loading.

A literature review has been carried out to obtain fatigue data for the target materials under loading conditions similar to the BDF target ones. Studies on  fatigue life are very limited for tungsten, TZM and tantalum (alloyed or un-alloyed), especially for fatigue behaviour at high temperatures and under irradiation. This makes the estimation of the fatigue properties for the BDF target materials difficult, as the different material and test conditions (geometry, heat treatment, purity, test mode, test stress ratio, test temperature, test frequency...) can considerably change the resulting fatigue strength value. 

Table~\ref{tab:TGT:fatiguedata} summarizes the values of fatigue data which are considered to represent the closest conditions to the BDF target operation. Nevertheless, it is worth mentioning that some of the test conditions are not identical to the ones of the BDF target materials (e.g. fatigue data given at room temperature, different material production route, etc.).

\begin{table*}[hbtp]
\begin{ruledtabular}

\caption{\label{tab:TGT:fatiguedata} Summary of the reviewed high-cycle fatigue data relevant for the BDF target operational conditions. Sources: TZM~\cite{TZMfatigue}, W~\cite{Wfatigue}, Ta2.5W~\cite{TantalumW2}. P/M: Powder Metallurgy, Aw: As worked, HIP: Hot Isostatic Pressing, Rxx: Recrystallized, RT: Room Temperature.}
{\renewcommand{\arraystretch}{1.2}%
\begin{tabular}{ccccccc}

{Material} & \begin{tabular}[c]{@{}c@{}}{Production} \\ {process}\end{tabular} & {Dimensions} & {Test mode} & \begin{tabular}[c]{@{}l@{}}{Number} \\ {of cycles}\end{tabular} & \begin{tabular}[c]{@{}c@{}} {Stress ratio,} \\ {temperature,} \\ {frequency (Hz)}\end{tabular} & \begin{tabular}[c]{@{}c@{}}{Fatigue limit}\\ {(MPa)}\end{tabular}\\
\hline
\multirow{2}{*}{{TZM}} & \multirow{2}{*}{P/M, Aw} & \multirow{2}{*}{\O50 mm bar} & \multirow{2}{*}{Push-pull} & \multirow{2}{*}{10\textsuperscript{7}} & -1, RT, 25 & {440} \\
 &  &  &  &  & -1, $850^{\circ}\mathrm{C}$, 25 & {250} \\
\hline
\multirow{2}{*}{{W}} & \begin{tabular}[c]{@{}c@{}}Sintered \\ + HIP\end{tabular} & \O5 mm bar  & Push-pull & $2\cdot10\textsuperscript{6}$ & 0, RT, 25 & {180}  \\
 & \begin{tabular}[c]{@{}c@{}}Rolled \\ + Annealed\end{tabular} & \O5 mm bar & Push-pull &  $2\cdot10\textsuperscript{6}$& 0, RT, 25 & {350}\\
\hline
{Ta2.5W} & P/M, Rxx & Plate 1 mm  & \begin{tabular}[c]{@{}c@{}}Bending \\ fatigue\end{tabular} & \begin{tabular}[c]{@{}c@{}}10\textsuperscript{7}, 50\% \\ fracture\end{tabular} & -1, RT, 25 & {310} \\
\end{tabular}}
\end{ruledtabular}
\end{table*}

It can be seen that TZM has the highest fatigue strength at $N=10\textsuperscript{7}$ with 440 MPa at room temperature, and the fatigue limit decreases at high temperatures as expected. Pure tungsten exhibits the lowest fatigue strength (180 MPa) for the sintered and HIPed manufacturing route, which is considered closer to the BDF target case. It is worth noting that the measured fatigue strength can be considered as conservative, since all the tests presented in Ref.~\cite{Wfatigue} were carried out in the tensile regime, and are assumed to lead to lower values than for fully-reversed loading. Ta2.5W shows a fatigue strength of 310 MPa at room temperature, which is relatively close to the tensile strength of the material (85\% of the UTS at room temperature). 

\subsubsection{Fatigue simulation results}

As a result of the high temperatures reached during operation for every beam impact on target, the target materials are subjected to large mean stresses and stress amplitudes. The fatigue data found in literature is usually obtained from uniaxial fully-reversed tests (with a mean stress equal to zero). However, the state of stresses in the target blocks is multiaxial and with a non-zero mean stress. In order to correlate the stresses calculated for the BDF target with the available fatigue strength, a two-step approach is necessary:

\begin{itemize}
    \item Firstly, an equivalent mean stress and an equivalent stress amplitude must be calculated, that are expected to give the same fatigue life in uniaxial loading as the multiaxial stress-state found in the target. 
    \item Secondly, an equivalent fully-reversed stress needs to be computed  from the values of equivalent mean stress and equivalent stress amplitude. This equivalent fully-reversed stress must take into account the contribution of the mean stress to the fatigue life of the target materials, in order to compare it with the fatigue strength under fully-reversed loading found in literature.
\end{itemize}

For TZM and Ta2.5W, which are considered as ductile materials, the Sines method~\cite{Fatemi} has been regarded as the most suited for this analysis. However, given the low availability of fatigue strength data under loading with non-zero mean stresses for the target materials, a similar approach requiring only data for fully-reversed fatigue has been adopted. 

The octaedral shear stress (von Mises) theory applied to a multiaxial state of stresses has been used to calculate the equivalent stress amplitude $\sigma_{q,a}$ and the equivalent mean stress $\sigma_{q,m}$ from the evolution of principal stresses in the target materials~\cite{Fatemi}. This equivalent stress method is limited to proportional loading conditions, where the principal directions remain unchanged during the loading cycle, which is the case for the BDF target blocks. It is worth noting that the value of $\sigma_{q,m}$ obtained from von Mises equation is always positive, and does not take into account the possible beneficial effects of compressive mean stresses. Given that the mean principal stresses of the target materials are compressive in most cases, the applied method can be regarded as conservative.

For pure tungsten, which is regarded as a brittle material, the maximum stress criterion has been considered~\cite{Fatemi}. It is expected that the fatigue endurance of the tungsten core under the multiaxial state of stresses during operation is principally influenced by the tensile contribution of the stress tensor. Therefore, the equivalent mean stress and the equivalent stress amplitude for pure tungsten have been taken from the maximum principal stress cyclic variation.

The equivalent mean stress and stress amplitude have been calculated for the three target materials, as shown in Table~\ref{tab:TGT:fatigueresults2}. The stress state represented by the equivalent mean stress and stress amplitude corresponds to uniaxial loading and is expected to be equivalent in terms of fatigue life to the actual multiaxial stress evolution in the target materials.

TZM displays the highest equivalent stress amplitude and mean stress, followed by Ta2.5W. The results displayed do not take into account the stress variations during the circular sweep of the beam presented in Figure~\ref{fig:TGT:stress_evolution_sweep}. The effect of the beam dilution on the fatigue life of the target materials shall be further studied in the future, but the calculations performed so far have shown that it is not expected to significantly affect the target lifetime. 

The modified Goodman equation (Equation~\ref{eq:TGT:Goodman}) has been used to compare the stresses calculated for the BDF target materials which have a non-zero mean stress, with the fully-reversed stress amplitude found in literature. This equation relates the equivalent mean stress $\sigma_{q,m}$ and equivalent stress amplitude $\sigma_{q,a}$ with an equivalent fully-reversed stress amplitude $\sigma_{q,f}$ which is assumed to give the same fatigue life:

\begin{equation}
\label{eq:TGT:Goodman}
\centering
\smallskip
\sigma _{q,f}=\frac{{\sigma _{q,a}}}{\left({1-\frac{\sigma_{q,m}}{\sigma_{UTS}}}\right)}\leq\sigma_{fat}\!\hspace{5pt} ,
\smallskip
\end{equation}

where $\sigma_{UTS}$ is the Ultimate Tensile Strength of the material, which has been taken for the present calculations from the measured values at room temperature cited in Section \ref{Sec:TGT:Simus:struct}, and $\sigma_{fat}$ is the fatigue limit of the material. The results of the calculations performed are summarized in Table~\ref{tab:TGT:fatigueresults2}.

\begin{table}[hbtp]
\begin{ruledtabular}
\caption{\label{tab:TGT:fatigueresults2} Equivalent mean stress (${\sigma_{q,\ m}}$), stress amplitude (${\sigma_{q,\ a}}$) and fully-reversed stress (${\sigma_{q,\ f}}$) calculated for the BDF target materials. Comparison with the fatigue strength of the materials for 10\textsuperscript{7} cycles~\cite{Wfatigue,TZMfatigue,TantalumW2}.}
{\renewcommand{\arraystretch}{1.2}%
\begin{tabular}{ccccccc}
{Material} & 
\begin{tabular}[c]{@{}c@{}}${\sigma_{q,\ m}}$\\ {(MPa)}\end{tabular} &
\begin{tabular}[c]{@{}c@{}}${\sigma_{q,\ a}}$\\ {(MPa)}\end{tabular} &
\begin{tabular}[c]{@{}c@{}}${\sigma_{UTS}}$\\ {(MPa)}\end{tabular} & \begin{tabular}[c]{@{}c@{}}${\sigma_{q,\ f}}$\\ {(MPa)}\end{tabular} & 
\begin{tabular}[c]{@{}c@{}}{Fatigue}\\{limit}\\ {(MPa)}\end{tabular} & \begin{tabular}[c]{@{}c@{}}{Safety}\\ {margin}\end{tabular} \\
\hline
{TZM} & 70 & 60 & 525 & {70} & 440 & 6.3\\
{W}  & 50 & 35 &570 & {40} & 180 & 4.5 \\
{Ta2.5W} & 50 & 45 &360 & {50} & 310 & 6.2\\
\end{tabular}}
\end{ruledtabular}
\end{table}

The stresses found in the BDF target are well within the endurance limits found in literature. The thermo-mechanical calculations performed have proven that it is unlikely that the target blocks will fail due to high-cycle fatigue. In this case the Ta2.5W cladding is not a critical aspect for the target operation, which is due to the fact that the fatigue limit of the material is relatively close to its tensile strength.

Further studies are required to evaluate the effects of radiation damage, as it could eventually reduce the fatigue life of the target, even though the safety margin is relatively large. Additional fatigue data under more representative loading conditions are necessary for an accurate estimation of the target materials fatigue life under operational conditions and will be performed in the future.

\section{Target cooling system}
\label{Sec:TGT:CoolingCFD}

\subsection{Cooling system design}
\label{Sec:TGT:CoolingCFD:design}

The target cooling system is one of the most critical parts of the BDF target design, given the high energy deposited on target during operation. Pressurized water has been chosen as cooling medium, other coolants such as air or helium would require a much higher flow rate to dissipate such a considerable amount of power; limited experience with gas cooled proton targets is one of the additional reasons. Several requirements have been considered for the cooling system design: first, a high water velocity is necessary to obtain an effective HTC between the cooling medium and the target blocks. Then, the pressure of the circuit should be high, and the pressure drop minimized, in order to ensure that the water in contact with the solid blocks is always below the boiling temperature. Furthermore, the cooling system design has been optimized in order to limit the increase of temperature in the circulating water, whilst minimizing the necessary flow rate.

\begin{figure*}
\resizebox{1\textwidth}{!}{
\includegraphics{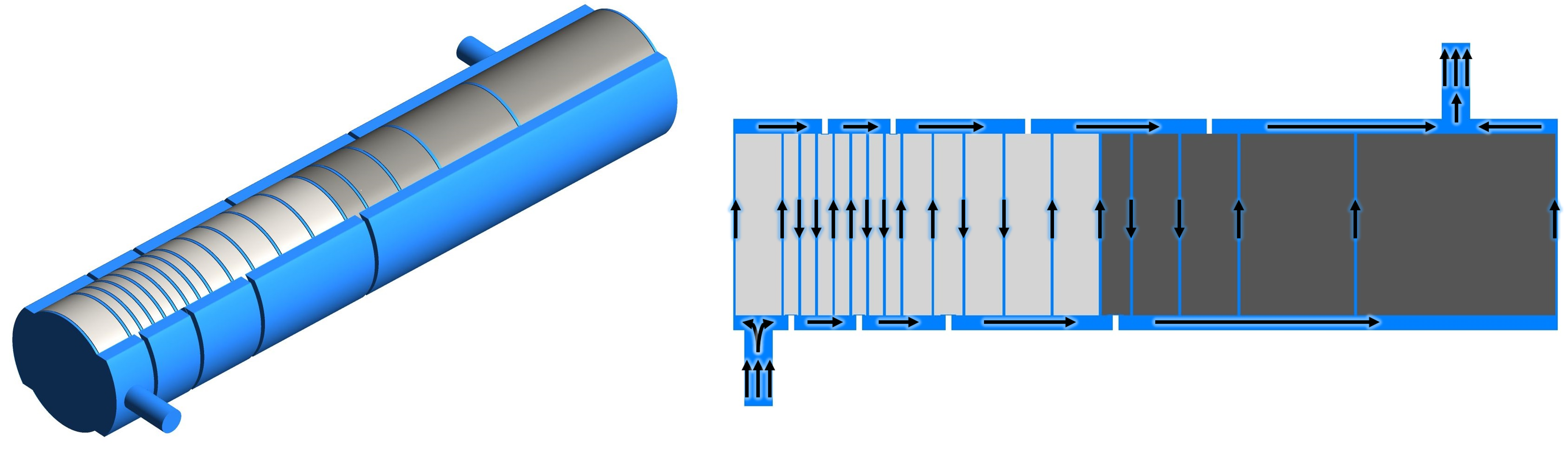}}
\caption{\label{fig:TGT:3Dcooling3}3D model of the BDF target cooling circuit and longitudinal cross-section (top view). The water flow is distributed by the manifolds (located in the left and right side of the target) to the 5 mm water channels between the target blocks. The serpentine configuration with two parallel streams (three for the last channels) can be observed.}
\end{figure*}

Figure~\ref{fig:TGT:3Dcooling3} illustrates the cooling system circulation path around the target blocks. The target cylinders are separated by 5 mm channels that allow water passage between the blocks. The water circulation is aimed at cooling down all the flat faces of the target cylinders, since they are impacted directly by the diluted beam and will therefore reach the highest temperatures during operation. The proposed cooling system design consists in a serpentine configuration (series flow) with two parallel streams. The serpentine circulation can provide high water speeds in the channels with a relatively low flow rate. 

The two-parallel-stream configuration aims at reducing the total pressure drop of the circuit and the temperature increase of the water from inlet to outlet. Another reason for this arrangement is to avoid a complete cooling circuit failure in the event that one of the channels is blocked (for example due to swelling of the blocks due to thermal expansion or debris in the circuit). If one of the channels is blocked, the water flow can continue through the other parallel channels, improving the circuit reliability. The last three channels are set in parallel, given that the number of channels is odd and the last tungsten blocks are the ones receiving the lowest amount of energy.

The channels are connected by "manifolds" that receive the water from two channels and distribute it to the following two. The manifold size is reduced to minimize the total water volume of the cooling circuit, and is constrained by the target support design, that will be described in Section~\ref{Sec:TGT:MechDesign}. The serpentine circulation is horizontal, from the left side of the blocks to the right side and vice versa; and is ensured by the presence of "blockers" after every two channels. This configuration has been chosen profiting from the experience of the target prototype cooling system design~\cite{sola2019beam}, where the series circulation of water was designed to be vertical (bottom-top-bottom), leading to the formation of air pockets during the filling process and to stagnant water in the bottom after drainage. 

A high pressure of 22 bar is selected to raise the water boiling point above 200$^\circ$C and avoid localized boiling during operation. The main risk of having localized boiling is the loss of heat transfer between the solid blocks and the water, which would prevent the heat dissipation by convection. 

It is desired to have a relatively uniform water velocity in the channels of around 5 m/s. Higher water speeds could lead to undesired erosion-corrosion effects on the Ta2.5W cladding surface. As will be shown in the following sections, the average HTC obtained with a velocity of 5 m/s in the channels is sufficiently high to ensure temperatures and stresses well within the operational limits of the target. No major gain is obtained by going to higher water speeds.

\subsection{Computational Fluid Dynamics calculations}
\label{Sec:TGT:CoolingCFD:CFD}

A commercial CFD code, ANSYS Fluent\copyright, has been used to perform the extensive 3D turbulence modelling of the flow configuration. A non-uniform energy deposition map imported from the FLUKA Monte Carlo~\cite{FLUKA_Code} simulation results has been applied to all the target blocks in the ANSYS Fluent model. In conjugate heat transfer problems (like the present one), the heat transmitted from the solid body to the liquid is highly dependent on the thermal boundary layer. Therefore, a sophisticated turbulence model, $k-\omega$ SST, has been used to resolve the boundary layers.

\subsubsection{Steady-state results}

Steady-state simulations have been carried out in order to investigate the flow behaviour in the cooling circuit and the target. The estimated pressure drop in the circuit is 3.2 bar, which means that the absolute pressure at the outlet of the target cooling circuit is around 19 bar. The boiling temperature at this pressure is over $200^{\circ} \text{C}$.

\begin{figure}
\resizebox{0.45\textwidth}{!}{
\includegraphics{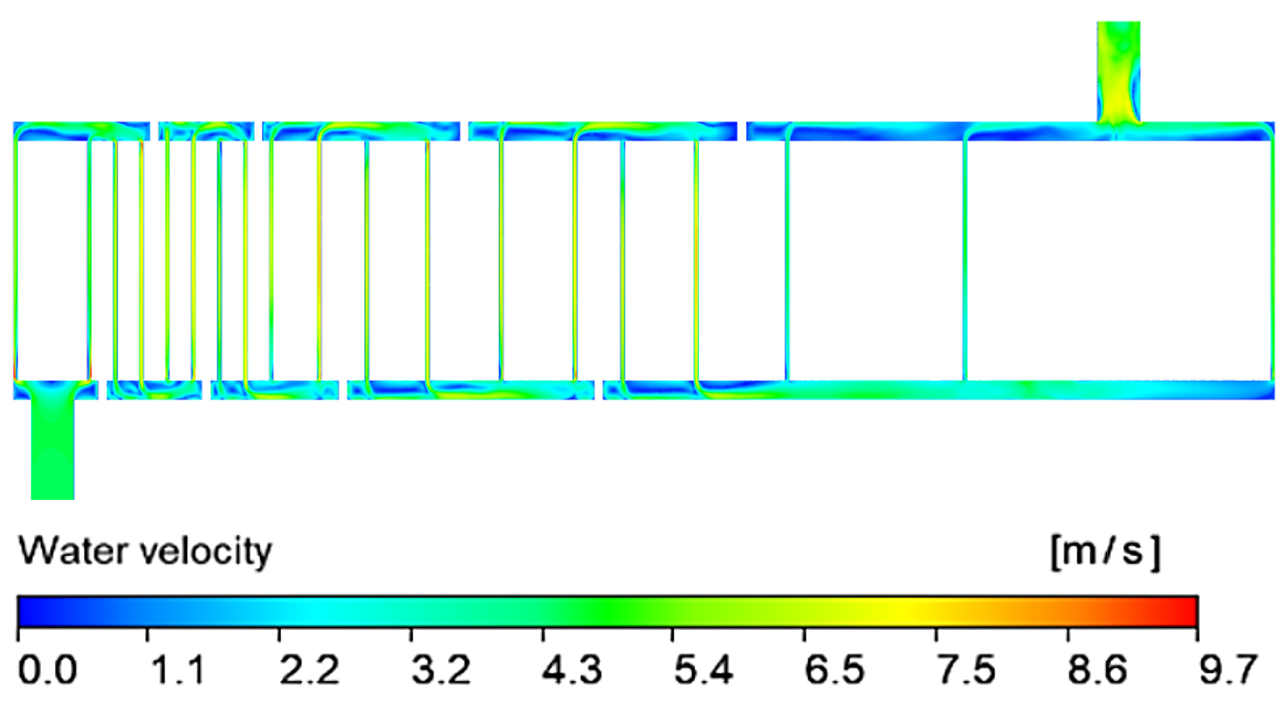}}
\caption{\label{fig:TGT:CFD_vel} 2-dimensional velocity contour in the target cooling circuit along the longitudinal plane. The water speed in all but the last three channels is relatively uniform around 5 m/s.}
\end{figure}

Figure~\ref{fig:TGT:CFD_vel} illustrates the velocity contour in the cooling domain along the longitudinal plane. It can be seen that the average velocity in all the cooling channels stands between 4 to 6 m/s, except for the last three channels where the average velocity is around 3 to 4 m/s. 

\begin{figure}
\resizebox{0.45\textwidth}{!}{
\includegraphics{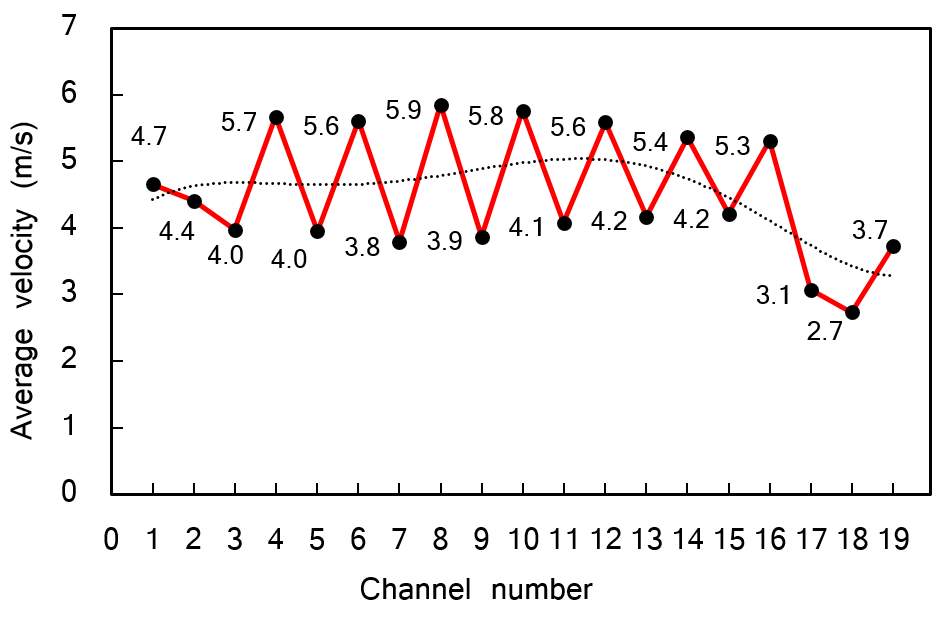}}
\caption{\label{fig:TGT:CFD_plot_velo} Variation of average velocities in the target water channels. The trendline of the plotted values is represented by a dotted line, and stands close to the desired water velocity of 5 m/s.}
\end{figure}

As the water stream enters the target cooling circuit from the inlet, the flow is split into two streams that enter the channels with similar mass flow rate. As a result, the average velocity in the first two cooling channels is almost identical. However, due to the presence of blockers (which make the serpentine design possible), the mass flow rate in the even channels is higher than the one of the odd channels. This twisting effect can be appreciated in Figure~\ref{fig:TGT:CFD_plot_velo}. 

The total temperature rise at the outlet obtained from the steady-state CFD simulations performed is around 8$^\circ$C, leading to an outlet temperature of 34$^\circ$C. This is found to be in good agreement with the temperature rise calculated from the energy balance. However, the transient calculations have shown that the outlet temperature is expected to present small fluctuations influenced by the beam impact on target.

The average heat transfer coefficient has been obtained for all the 19 channels of the cooling system. The average HTC stands slightly above 20000 W/(m$^{2}\:$K) in all channels, except for the last three where lower velocities are found. This value is consistent with the HTC predicted by means of analytical calculations, which estimated an average value of 22000 W/(m$^{2}\:$K) in the cooling channels, as a result of the high water speed of around 5 m/s.

\subsubsection{Transient results}

CFD transient calculations have been performed to reproduce the target behaviour during the one second SPS primary beam impact on target and the subsequent 6.2 seconds cooling, thereby replicating the 7.2 seconds BDF cycle. Unlike in the thermal simulation setup shown in Section~\ref{Sec:TGT:Simus:thermal}, the time-dependent beam sweep over the one second spill is not taken into account, instead a circle-shape energy distribution corresponding to the circular dilution path is directly applied on the target blocks.

\begin{figure}
\resizebox{0.45\textwidth}{!}{
\includegraphics{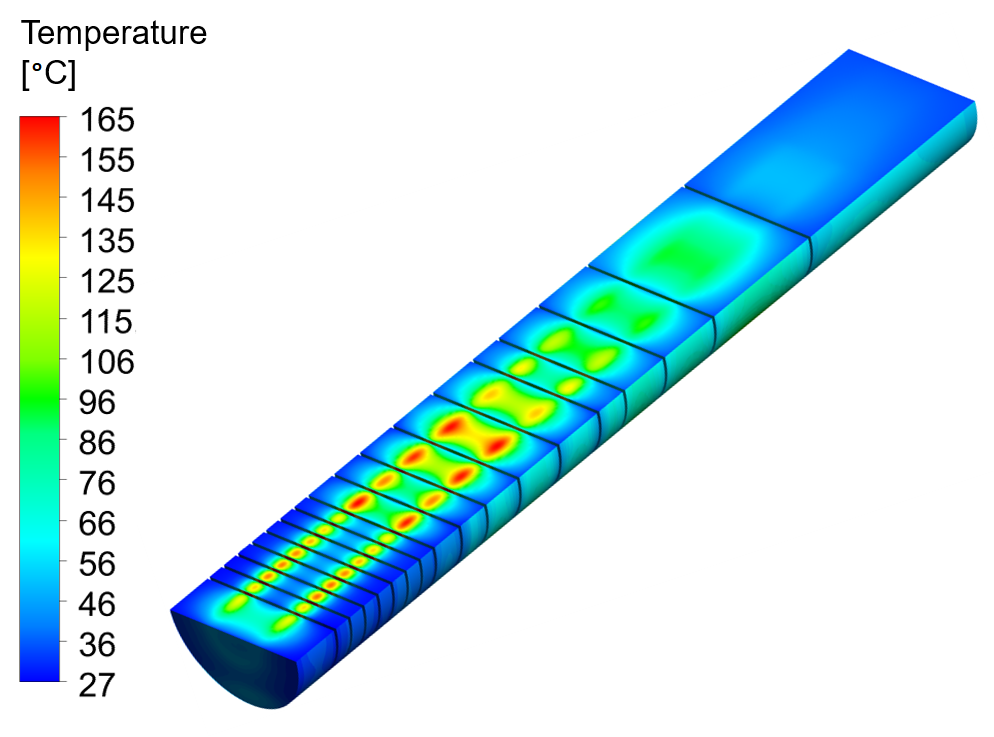}}
\caption{\label{fig:TGT:CFD_3D_temp} Half-model temperature distribution in the target blocks at the end of the proton pulse (1 second) during steady state conditions, with the HTC evaluated from CFD simulations. The maximum temperature expected is around 160$^{\circ}\text{C}$, found in blocks 9 and 12.}
\end{figure}

\begin{figure}
\resizebox{0.45\textwidth}{!}{
\includegraphics{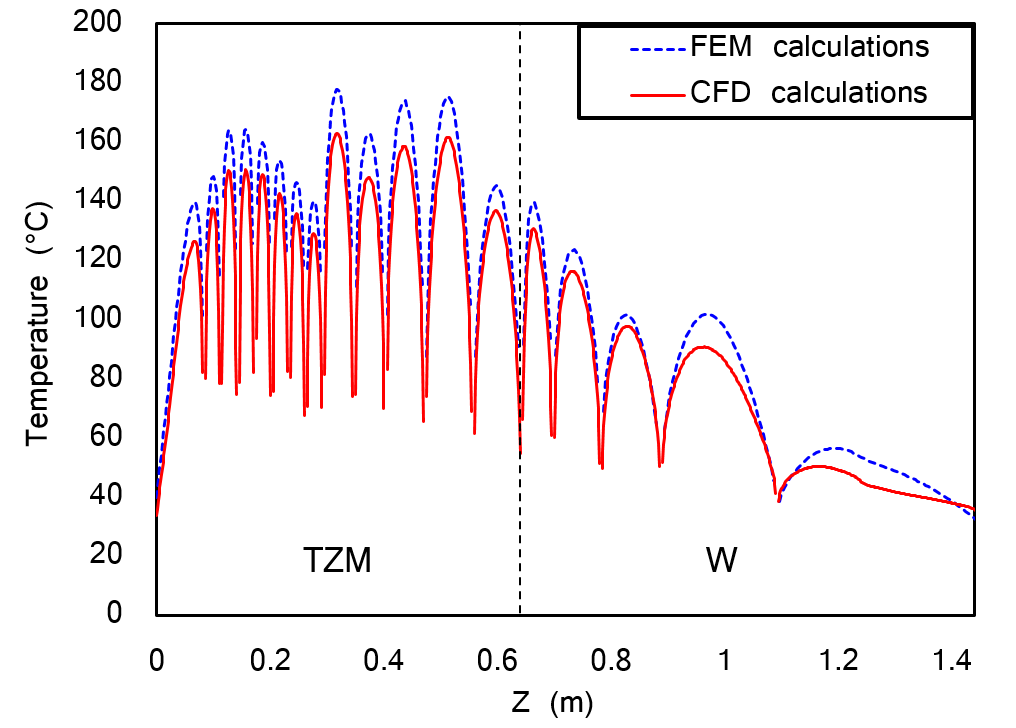}}
\caption{\label{fig:TGT:CFD_temp_transient} Temperature profile following a line parallel to the Z-axis and at a distance of 50 mm from the target axis (coincident with the beam impact location). Comparison between FEM and CFD calculations, results obtained at the end of the proton pulse during steady state conditions.}
\end{figure}

Figure~\ref{fig:TGT:CFD_3D_temp} illustrates the temperature distribution contour in a longitudinal section through all the target blocks after one SPS pulse on target. The corresponding temperature profile in the different blocks is plotted in Figure~\ref{fig:TGT:CFD_temp_transient}. Figure~\ref{fig:TGT:CFD_temp_transient} also shows the results for the identical time and location obtained from the FEM calculations assuming a constant HTC value.

The maximum temperatures in both cases are found in the TZM core of blocks 9 and 12, with temperatures around 160$^{\circ}\text{C}$ for the CFD calculations. In the tungsten core, the maximum temperatures are located in the first tungsten block, reaching around 130$^{\circ}\text{C}$. The temperature reached in the Ta2.5W cladding, which is considered one of the most critical elements for the target design, is about 120$^\circ$C for the CFD calculations, and is found in the cladding of block 3.

The temperature profile obtained for the FEM and CFD simulations is qualitatively similar, but the maximum temperatures expected from the CFD simulations are lower. This difference can be explained by the higher accuracy in the calculation of the HTC for the CFD simulations compared with the constant convection coefficient employed in the FEM analysis. It can be concluded that the structural stresses calculated in Section~\ref{Sec:TGT:Simus:struct} by importing the temperatures obtained via FEM calculations are conservative, and the safety margins with respect to the material limits are expected to be even higher. 
 
The surface temperature has been calculated for each target block, reaching a maximum temperature of around 90$^{\circ}$C in the upstream surface of block \#3. This value is less than half of the boiling temperature of water at the outlet pressure (above 210$^{\circ}$C), thus it can be concluded that water boiling is not likely to occur during normal operation of the BDF target.

\subsection{Considerations about loss of coolant accidents}

The target cooling circuit is critical for the successful operation of the BDF target. A disconnection or rupture of the cooling pipes or a failure of the cooling system equipment could lead to the sudden stop of the water circulation or to the loss of cooling water in the circuit. In the event of a Loss Of Coolant Accident (LOCA) with beam stop, the heat produced by the decay of the long-time irradiated target materials will dissipate at a very slow rate, due to the absence of forced cooling, increasing the temperature of the target materials for the hours and days following the accident. The main risks are the possible melting of the target materials given the high concentration of heat in the target or the production of tungsten trioxide (WO$_3$) species above 700$^{\circ}$C), which are highly volatile~\cite{WO3_oxi,Garoby_2017}.

\begin{figure}
\resizebox{0.45\textwidth}{!}{
\includegraphics{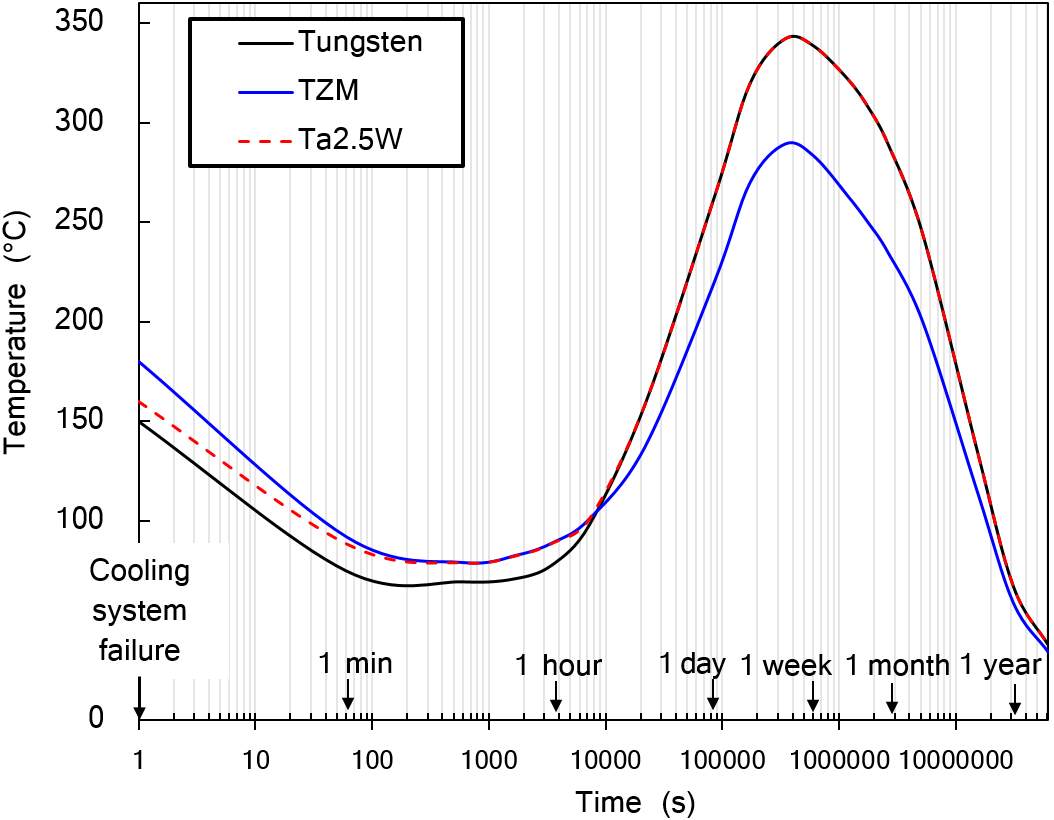}}
\caption{\label{fig:TGT:Decay_temp_2y} Evolution of the maximum temperature in the different target materials after a cooling system failure due to the decay heat generated inside the target (logarithmic scale).}
\end{figure}

The decay heat produced in the target materials for different periods has been estimated by FLUKA simulations, and thermal calculations have been carried out to evaluate the effects on target of a cooling system failure. A low convection coefficient of 1 W/(m$^{2}$K) has been selected as boundary condition for the calculations. Such a film convection coefficient could be considered as conservative for natural convection with air, helium or water. Figure~\ref{fig:TGT:Decay_temp_2y} presents the maximum temperature evolution in the target blocks during two years after a cooling system failure with immediate beam stop. Decay heat measurements have been performed for the neutron spallation target of ISIS (RAL, UK), and a similar trend has been reported~\cite{Decay_RAL}. The maximum temperature in the target blocks is around 350$^{\circ}$C and is reached one week after a cooling system failure; this is sufficiently low to avoid melting of the target materials and oxidation even in the case of breached tungsten. Further studies are required in order to understand the possibility for producing oxidation at these temperatures in TZM or Ta2.5W.

 The stresses induced by the thermal loads have also been calculated, concluding that the level of stresses is not critical in terms of fracture of the target materials, but could induce plastic deformation in the Ta2.5W cladding of some of the blocks. In order to achieve an accurate modelling of the air, helium or water behavior after a cooling system failure accident, CFD studies applied to this specific case are necessary.

\section{Target assembly mechanical design}
\label{Sec:TGT:MechDesign}

\begin{figure*}
\resizebox{0.88\textwidth}{!}{
\includegraphics{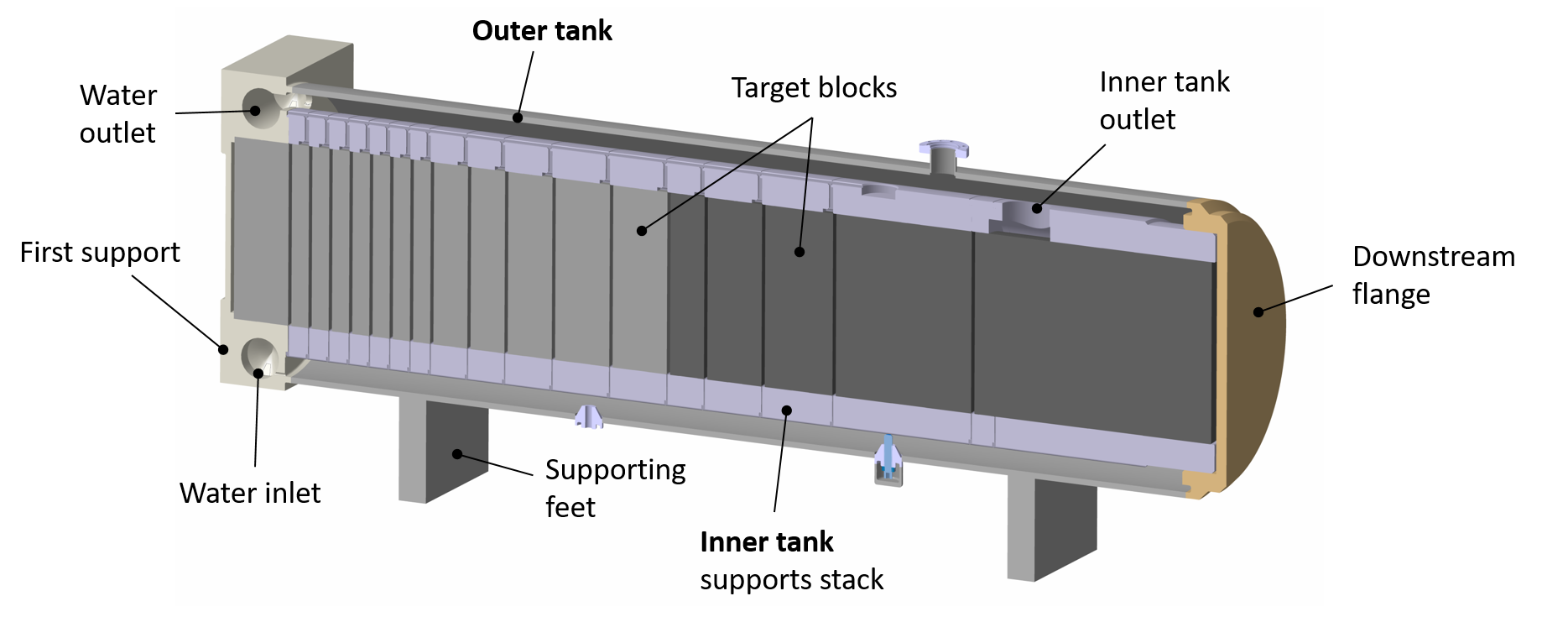}}
\caption{\label{fig:TGT:assembly_cut} Longitudinal section of the BDF target inner and outer tank structure. The target blocks supported by the inner tank can be seen, as well as many functional elements of the cooling circuit and the supporting structure of the target itself.}
\end{figure*}

\begin{figure}
\resizebox{0.48\textwidth}{!}{
\includegraphics{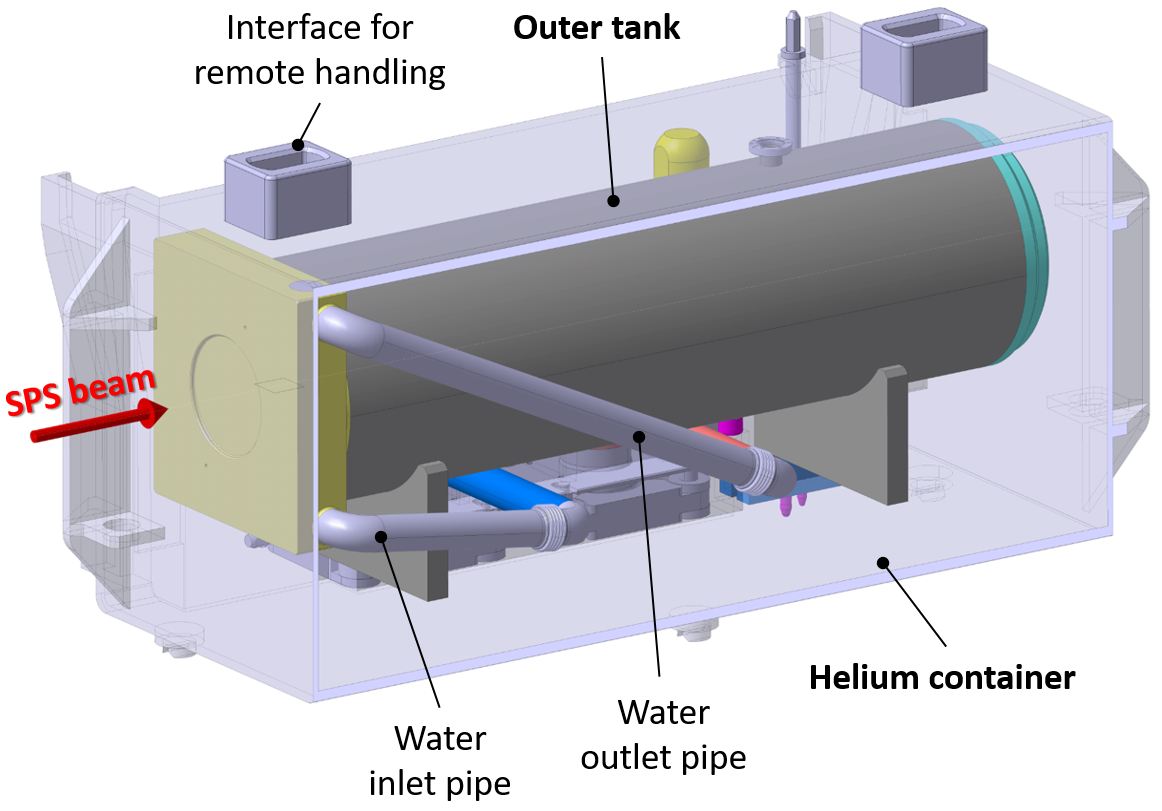}}
\caption{\label{fig:TGT:He_box} BDF target full assembly: view of the Helium container, outer tank, upstream and downstream flanges and inlet and outlet pipes.}
\end{figure}

The BDF target assembly consists of four main parts: the target blocks, an inner tank that supports the target blocks, a leak-tight outer tank that encloses the inner tank, and a helium container that contains the whole assembly. A section of the inner and outer tank supporting the target blocks can be seen in Figure~\ref{fig:TGT:assembly_cut}. The helium container enclosing the full target assembly is shown in Figure~\ref{fig:TGT:He_box}.

\subsection{Target blocks production}

The BDF target core blocks consist of two different parts:
\begin{itemize}
    \item A TZM or W cylinder with different length according to the block position in the target core. Early investigations have shown that all the TZM cylinders can be manufactured via multi-axial forging, while not all the pure tungsten cylinders can be obtained by this method~\cite{Production_plansee}. The length of some of the tungsten cylinders, that reach up to 350 mm long, is a limiting factor to apply longitudinal forging. For that reason, it is foreseen to produce the W cylinders via sintering and HIPing, this process leading to an isotropic material structure and an acceptable density of around 97\%. 
    \item A cladding made out of Ta2.5W, which encloses the TZM or W cylinder, and consists of a tube with variable length and two disks. The Ta2.5W tubes can be rolled, and must be seamless as this is a requirement for the HIP process that will be described later on. The Ta2.5W disks can be obtained by forging.
\end{itemize}

For the production of the target blocks, the TZM or W cylinder is inserted into the Ta2.5W tube and closed above and below by the two Ta2.5W disks. The top and bottom Ta2.5W disks are electron-beam (EB) welded to the Ta2.5W tube, and the capsules are covered with an Zirconium foil to prevent oxidation. Then, every assembled target block undergoes a HIP cycle, reaching a temperature of $1200\,^{\circ}\mathrm{C}$ and a pressure of 150 MPa for 2 hours. The HIP process carried out for the BDF target blocks production is crucial to ensure the mechanical and chemical bonding between the cladding and core materials. More details about the HIP assisted diffusion bonding carried out for the target blocks production are given in Ref.~\cite{HIP_Busom}.

\subsection{Target vessel inner stainless steel tank}

\begin{figure}
\resizebox{0.45\textwidth}{!}{
\includegraphics{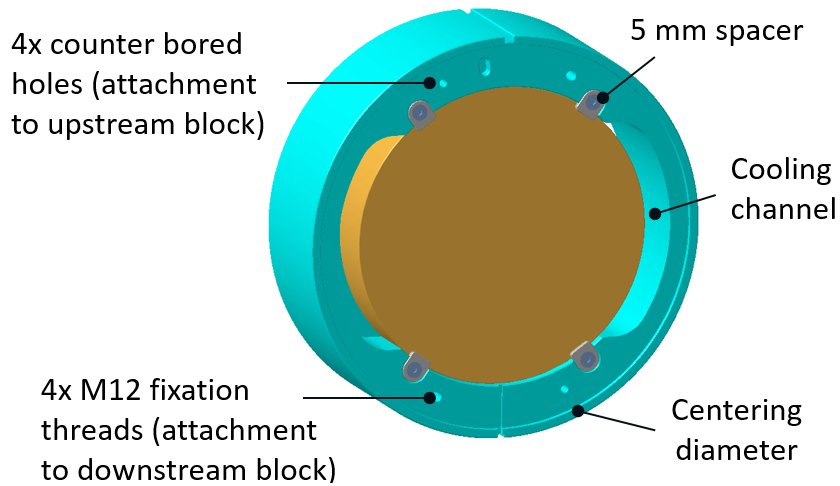}}
\caption{\label{fig:TGT:support_description} Description of one target block support, part of the target inner tank. Each support integrates different elements that permit the assembly with the previous and following supports, as well as the water circulation around the target blocks.}
\end{figure}

The inner tank is composed of several "supports" that are assembled together. Each target block has its own support, that is acting at the same time as a handling tool. Figure~\ref{fig:TGT:support_description} shows a description of the supports that make up the inner tank. 

The target blocks can weigh up to 300 kg for the heaviest tungsten cylinder, making it necessary to have a handling mechanism for their assembly. Each support holds the corresponding target block in a vertical position during assembly, and all the supports are stacked on top of each other starting from the first support. Once the assembly is completed, the target blocks supporting structure (hereafter, inner tank) is placed in horizontal position for the subsequent operation, as illustrated in Figure~\ref{fig:TGT:assembly_cut}.

Another function of the inner tank is to enclose the target cooling circuit. The different supports include dedicated grooves in order to provide the foreseen circulation path for the water cooling, compatible with the cooling system design presented in Section~\ref{Sec:TGT:CoolingCFD}. The 5 mm spacers added to the supports to ensure the water circulation between the blocks (Figure~\ref{fig:TGT:support_description}) also allow the blocks to be held in vertical position for the handling process.

\subsection{Target vessel outer stainless steel tank}

The outer tank is responsible for providing leak-tightness to the target assembly, and structural stability to the equipment. The cooling circuit is enclosed by the inner tank, that is - by design - not sealed against water leaks. The outer tank is welded to the first support at one end, and to the downstream flange at the other end (Figure~\ref{fig:TGT:assembly_cut}). The fact of having only two welds adds simplicity and robustness to the target manufacturing and assembly process, while ensuring a good reliability of the system. 

In order to avoid any stagnation of water between the inner and outer tank, a sufficiently large gap is foreseen to force the water circulation in the volume between both tanks, creating an ``external'' cooling loop, which can be seen in Figure~\ref{fig:TGT:assembly_cut}. The outer tank diameter has been optimized via CFD calculations, minimizing the pressure drop in the external circuit while keeping an acceptable water speed.

\begin{figure}
\resizebox{0.45\textwidth}{!}{
\includegraphics{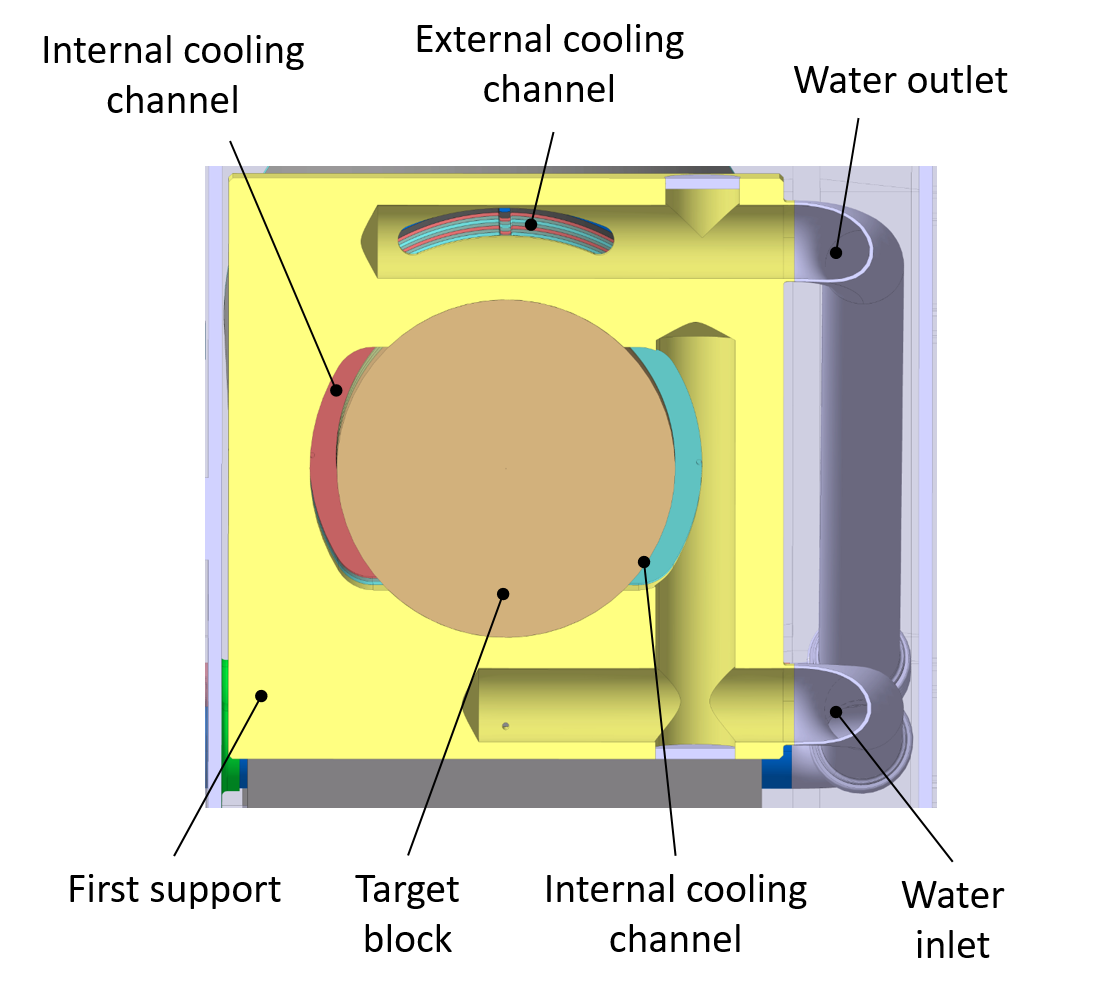}}
\caption{\label{fig:TGT:First_support} Layout of the first support of the target inner tank, made out of stainless steel 316L, also functioning as upstream flange of the outer tank and beam window.}
\end{figure}

Figure~\ref{fig:TGT:First_support} shows the layout of the first support of the inner tank, one of the key elements of the target assembly. This support acts as upstream flange of the outer tank, and includes both the internal cooling circuit inlet and the external cooling circuit outlet. 

\begin{figure}
\resizebox{0.4\textwidth}{!}{
\includegraphics{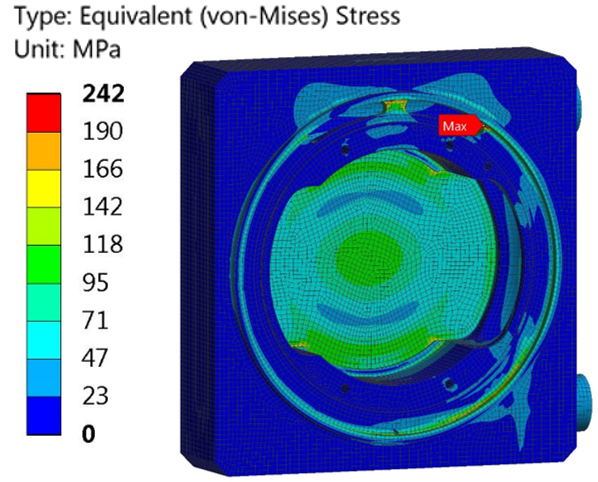}}
\caption{\label{fig:TGT:First_support_stress} Maximum stress distribution in the first support and beam window at an over-pressure of 32 bar, calculated by means of FEM simulations. The maximum Von Mises equivalent stress in the beam window is around 100 MPa.}
\end{figure}

The first support assembly includes the proton beam window. The thickness of the central part of the first support, where the diluted beam impact will take place, has been optimized to obtain a beam window thick enough for good mechanical reliability, taking into account the internal operational pressure of 22 bar; whilst being thin enough to avoid critical levels of energy deposition due to the beam impact. As shown in Figure~\ref{fig:TGT:First_support_stress}, the maximum stress expected in the outer tank beam window at an over-pressure of 32 bar is about 100 MPa, well below the yield strength of stainless steel under the operational conditions (estimated at around 200 MPa). The temperature increase due to the beam interaction with the proton beam window was estimated through FEM simulations and is below 5\,$^{\circ}\mathrm{C}$ per pulse. Consequently, the thermally-induced stresses are very low and were considered to be negligible.

\subsection{Target helium tank}

The whole target assembly is contained inside a square-section tank filled with inert gas (He at the current stage), as shown in Figure~\ref{fig:TGT:He_box}. The presence of helium gas ensures a dry and controlled environment for the target operation, reducing the corrosion effects on the target assembly components. Additionally, the closed circulation of helium allows the monitoring of possible water leaks from the target vessel. 

In case of target failure, it is foreseen to replace the whole helium container with its internal components. A comprehensive study of the target complex handling and integration has been performed and reported in detail in a separate publication~\cite{BDFcomplex}. The target assembly design is fully compatible with the target complex integration, that foresees the target helium container remote disconnection and exchange. All the electrical, water and helium interfaces of the helium tank that can be seen in Figure~\ref{fig:TGT:He_box} have been designed to be disconnected via remote handling techniques, and are detailed in~\cite{BDFcomplex}.

\section{Conclusions and future work}
In the framework of the Beam Dump Facility comprehensive design study phase, the design of the target system has been developed. Based on the experimental requirements of the facility, an innovative solution consisting of a dense target/dump made of several blocks of refractory metals clad with a thin layer of tantalum-alloy has been developed. The proposed design is capable of dealing with the average beam power of 355 kW and with a deposited energy of 2.56 MJ per pulse every 6.2 seconds. Due to the beam energy and resulting temperature and stress profiles, new materials have been introduced, including TZM for the core and Ta2.5W for the cladding.

A target composed of several collinear cylinders segmented in an optimized manner is foreseen, and a circular beam dilution system has been selected in order to deal with the challenging levels of heat and energy deposition induced by the SPS proton beam interaction. Extensive thermo-mechanical studies have been carried out to evaluate the target reliability at the high temperatures, stresses and high-cycle loading to which the target will be exposed during its lifetime. The most critical element in the target assembly is the Ta2.5W cladding, which is expected to reach temperatures close to 200$^\circ$C and cyclic stresses of around 100 MPa. The stresses and temperatures obtained from the thermal and structural analysis performed are within the material limits, providing reassurance on the soundness of the proposed target design and material selection. As heat dissipation is a key factor in the target survivability, a high-speed water cooling system is foreseen; the proposed system will lead to effective heat removal from the target. A mechanical design of the current version of target assembly has been executed in order to evaluate the feasibility of the manufacturing, handling and assembly processes. Finally, a target prototype has been built and successfully tested under a 40 kW proton beam for few 10$^{16}$ protons on target; results will be reported in details elsewhere, but the tests indicated a very good agreement between the experimentally measured temperatures and stresses and the simulations, providing confidence on the developed target model. 

The studies described in the present contribution constitute an assessment of the first generation Beam Dump Facility target, and further studies are already foreseen in the framework of the facility Technical Design phase. 
  
\section{Acknowledgments}

The authors express their thanks to the Physics Beyond Colliders Study Group (specifically the Beam Dump Facility Project group) for the support in the execution of the activity. The authors would also like to thank Dan Wilcox (ISIS/RAL, United Kingdom) for technical exchanges and valuable discussions.

\bibliography{references}

\begin{thebibliography}{45}%
\makeatletter
\providecommand \@ifxundefined [1]{%
 \@ifx{#1\undefined}
}%
\providecommand \@ifnum [1]{%
 \ifnum #1\expandafter \@firstoftwo
 \else \expandafter \@secondoftwo
 \fi
}%
\providecommand \@ifx [1]{%
 \ifx #1\expandafter \@firstoftwo
 \else \expandafter \@secondoftwo
 \fi
}%
\providecommand \natexlab [1]{#1}%
\providecommand \enquote  [1]{``#1''}%
\providecommand \bibnamefont  [1]{#1}%
\providecommand \bibfnamefont [1]{#1}%
\providecommand \citenamefont [1]{#1}%
\providecommand \href@noop [0]{\@secondoftwo}%
\providecommand \href [0]{\begingroup \@sanitize@url \@href}%
\providecommand \@href[1]{\@@startlink{#1}\@@href}%
\providecommand \@@href[1]{\endgroup#1\@@endlink}%
\providecommand \@sanitize@url [0]{\catcode `\\12\catcode `\$12\catcode
  `\&12\catcode `\#12\catcode `\^12\catcode `\_12\catcode `\%12\relax}%
\providecommand \@@startlink[1]{}%
\providecommand \@@endlink[0]{}%
\providecommand \url  [0]{\begingroup\@sanitize@url \@url }%
\providecommand \@url [1]{\endgroup\@href {#1}{\urlprefix }}%
\providecommand \urlprefix  [0]{URL }%
\providecommand \Eprint [0]{\href }%
\providecommand \doibase [0]{http://dx.doi.org/}%
\providecommand \selectlanguage [0]{\@gobble}%
\providecommand \bibinfo  [0]{\@secondoftwo}%
\providecommand \bibfield  [0]{\@secondoftwo}%
\providecommand \translation [1]{[#1]}%
\providecommand \BibitemOpen [0]{}%
\providecommand \bibitemStop [0]{}%
\providecommand \bibitemNoStop [0]{.\EOS\space}%
\providecommand \EOS [0]{\spacefactor3000\relax}%
\providecommand \BibitemShut  [1]{\csname bibitem#1\endcsname}%
\let\auto@bib@innerbib\@empty
\bibitem [{\citenamefont {Bonivento}\ \emph {et~al.}(2013)\citenamefont
  {Bonivento} \emph {et~al.}}]{EOI_SHiP}%
  \BibitemOpen
  \bibfield  {author} {\bibinfo {author} {\bibfnamefont {W.}~\bibnamefont
  {Bonivento}} \emph {et~al.},\ }\bibfield  {title} {\enquote {\bibinfo {title}
  {{Proposal to Search for Heavy Neutral Leptons at the SPS}},}\ }\href@noop {}
  {\  (\bibinfo {year} {2013})},\ \Eprint {http://arxiv.org/abs/1310.1762}
  {arXiv:1310.1762 [hep-ex]} \BibitemShut {NoStop}%
\bibitem [{\citenamefont {Anelli}\ \emph {et~al.}(2015)\citenamefont {Anelli}
  \emph {et~al.}}]{Anelli_SHiP}%
  \BibitemOpen
  \bibfield  {author} {\bibinfo {author} {\bibfnamefont {M.}~\bibnamefont
  {Anelli}} \emph {et~al.} (\bibinfo {collaboration} {SHiP Collaboration}),\
  }\bibfield  {title} {\enquote {\bibinfo {title} {{A facility to Search for
  Hidden Particles (SHiP) at the CERN SPS}},}\ }\href@noop {} {\  (\bibinfo
  {year} {2015})},\ \Eprint {http://arxiv.org/abs/1504.04956} {arXiv:1504.04956
  [physics.ins-det]} \BibitemShut {NoStop}%
\bibitem [{\citenamefont {{S. Alekhin \textit{et al.} (SHiP
  Collaboration)}}(2016)}]{Alekhin_SHiP}%
  \BibitemOpen
  \bibfield  {author} {\bibinfo {author} {\bibnamefont {{S. Alekhin \textit{et
  al.} (SHiP Collaboration)}}},\ }\bibfield  {title} {\enquote {\bibinfo
  {title} {{A facility to Search for Hidden Particles at the CERN SPS: the SHiP
  physics case}},}\ }\href {\doibase 10.1088/0034-4885/79/12/124201} {\bibfield
   {journal} {\bibinfo  {journal} {Rept. Prog. Phys.}\ }\textbf {\bibinfo
  {volume} {79}},\ \bibinfo {pages} {124201} (\bibinfo {year} {2016})},\
  \Eprint {http://arxiv.org/abs/1504.04855} {arXiv:1504.04855 [hep-ph]}
  \BibitemShut {NoStop}%
\bibitem [{\citenamefont {\textit{et al.}}(2019{\natexlab{a}})}]{Ahdida_2019}%
  \BibitemOpen
  \bibfield  {author} {\bibinfo {author} {\bibfnamefont {C.~Ahdida}\
  \bibnamefont {\textit{et al.}}},\ }\bibfield  {title} {\enquote {\bibinfo
  {title} {The experimental facility for the search for hidden particles at the
  {CERN} {SPS}},}\ }\href {\doibase 10.1088/1748-0221/14/03/p03025} {\bibfield
  {journal} {\bibinfo  {journal} {Journal of Instrumentation}\ }\textbf
  {\bibinfo {volume} {14}},\ \bibinfo {pages} {P03025--P03025} (\bibinfo {year}
  {2019}{\natexlab{a}})}\BibitemShut {NoStop}%
\bibitem [{\citenamefont {Kershaw}\ and\ \citenamefont {\textit{et
  al.}}(2018)}]{BDFcomplex}%
  \BibitemOpen
  \bibfield  {author} {\bibinfo {author} {\bibfnamefont {K.}~\bibnamefont
  {Kershaw}}\ and\ \bibinfo {author} {\bibfnamefont {M.~Calviani}\ \bibnamefont
  {\textit{et al.}}},\ }\bibfield  {title} {\enquote {\bibinfo {title} {Design
  development for the {Beam Dump Facility Target Complex} at {CERN}},}\ }\href
  {\doibase 10.1088/1748-0221/13/10/p10011} {\bibfield  {journal} {\bibinfo
  {journal} {Journal of Instrumentation}\ }\textbf {\bibinfo {volume} {13}},\
  \bibinfo {pages} {P10011--P10011} (\bibinfo {year} {2018})}\BibitemShut
  {NoStop}%
\bibitem [{\citenamefont {Thomason}(2019)}]{THOMASON201961}%
  \BibitemOpen
  \bibfield  {author} {\bibinfo {author} {\bibfnamefont {J.W.G.}\ \bibnamefont
  {Thomason}},\ }\bibfield  {title} {\enquote {\bibinfo {title} {{The ISIS
  Spallation Neutron and Muon Source: The first thirty-three years}},}\ }\href
  {\doibase https://doi.org/10.1016/j.nima.2018.11.129} {\bibfield  {journal}
  {\bibinfo  {journal} {Nuclear Instruments and Methods in Physics Research
  Section A: Accelerators, Spectrometers, Detectors and Associated Equipment}\
  }\textbf {\bibinfo {volume} {917}},\ \bibinfo {pages} {61 -- 67} (\bibinfo
  {year} {2019})}\BibitemShut {NoStop}%
\bibitem [{\citenamefont {Bungau}\ \emph {et~al.}(2014)\citenamefont {Bungau},
  \citenamefont {Bungau}, \citenamefont {Cywinski},\ and\ \citenamefont
  {Edgecock}}]{Bungau:2014nda}%
  \BibitemOpen
  \bibfield  {author} {\bibinfo {author} {\bibfnamefont {C.}~\bibnamefont
  {Bungau}}, \bibinfo {author} {\bibfnamefont {A.}~\bibnamefont {Bungau}},
  \bibinfo {author} {\bibfnamefont {R.}~\bibnamefont {Cywinski}}, \ and\
  \bibinfo {author} {\bibfnamefont {T.}~\bibnamefont {Edgecock}},\ }\bibfield
  {title} {\enquote {\bibinfo {title} {{Power Upgrade Studies for the ISIS-TS1
  Spallation Target}},}\ }in\ \href {\doibase 10.18429/JACoW-IPAC2014-THPRI082}
  {\emph {\bibinfo {booktitle} {{Proceedings, 5th International Particle
  Accelerator Conference (IPAC 2014): Dresden, Germany, June 15-20, 2014}}}}\
  (\bibinfo {year} {2014})\ p.\ \bibinfo {pages} {THPRI082}\BibitemShut
  {NoStop}%
\bibitem [{\citenamefont {Dey}\ and\ \citenamefont {Jones}(2018)}]{DEY201863}%
  \BibitemOpen
  \bibfield  {author} {\bibinfo {author} {\bibfnamefont {A.}~\bibnamefont
  {Dey}}\ and\ \bibinfo {author} {\bibfnamefont {L.}~\bibnamefont {Jones}},\
  }\bibfield  {title} {\enquote {\bibinfo {title} {{Strategies to improve ISIS
  TS2 target life}},}\ }\href {\doibase
  https://doi.org/10.1016/j.jnucmat.2017.12.044} {\bibfield  {journal}
  {\bibinfo  {journal} {Journal of Nuclear Materials}\ }\textbf {\bibinfo
  {volume} {506}},\ \bibinfo {pages} {63 -- 70} (\bibinfo {year} {2018})},\
  \bibinfo {note} {special Issue on Spallation Materials Technology – Select
  Papers from the Thirteenth International Workshop on Spallation Materials
  Technology (IWSMT-13)}\BibitemShut {NoStop}%
\bibitem [{\citenamefont {Nowicki}\ \emph {et~al.}(2017)\citenamefont
  {Nowicki}, \citenamefont {Wender},\ and\ \citenamefont
  {Mocko}}]{NOWICKI2017374}%
  \BibitemOpen
  \bibfield  {author} {\bibinfo {author} {\bibfnamefont {S.~F.}\ \bibnamefont
  {Nowicki}}, \bibinfo {author} {\bibfnamefont {S.~A.}\ \bibnamefont {Wender}},
  \ and\ \bibinfo {author} {\bibfnamefont {M.}~\bibnamefont {Mocko}},\
  }\bibfield  {title} {\enquote {\bibinfo {title} {{The Los Alamos Neutron
  Science Center Spallation Neutron Sources}},}\ }\href {\doibase
  https://doi.org/10.1016/j.phpro.2017.09.035} {\bibfield  {journal} {\bibinfo
  {journal} {Physics Procedia}\ }\textbf {\bibinfo {volume} {90}},\ \bibinfo
  {pages} {374 -- 380} (\bibinfo {year} {2017})},\ \bibinfo {note} {conference
  on the Application of Accelerators in Research and Industry, CAARI 2016, 30
  October – 4 November 2016, Ft. Worth, TX, USA}\BibitemShut {NoStop}%
\bibitem [{\citenamefont {\textit{et al.}}(2008)}]{PhysRevD.78.052002}%
  \BibitemOpen
  \bibfield  {author} {\bibinfo {author} {\bibfnamefont {K.~Kodama}\
  \bibnamefont {\textit{et al.}}} (\bibinfo {collaboration} {DONuT
  Collaboration}),\ }\bibfield  {title} {\enquote {\bibinfo {title} {{Final
  tau-neutrino results from the DONuT experiment}},}\ }\href {\doibase
  10.1103/PhysRevD.78.052002} {\bibfield  {journal} {\bibinfo  {journal} {Phys.
  Rev. D}\ }\textbf {\bibinfo {volume} {78}},\ \bibinfo {pages} {052002}
  (\bibinfo {year} {2008})}\BibitemShut {NoStop}%
\bibitem [{\citenamefont {\textit{et al.}}(1986)}]{DORENBOSCH1986473}%
  \BibitemOpen
  \bibfield  {author} {\bibinfo {author} {\bibfnamefont {J.~Dorenbosch}\
  \bibnamefont {\textit{et al.}}},\ }\bibfield  {title} {\enquote {\bibinfo
  {title} {{A search for decays of heavy neutrinos in the mass range 0.5–2.8
  GeV}},}\ }\href {\doibase https://doi.org/10.1016/0370-2693(86)91601-1}
  {\bibfield  {journal} {\bibinfo  {journal} {Physics Letters B}\ }\textbf
  {\bibinfo {volume} {166}},\ \bibinfo {pages} {473 -- 478} (\bibinfo {year}
  {1986})}\BibitemShut {NoStop}%
\bibitem [{\citenamefont
  {Bergsma}(1988)}]{doi:10.1111/j.1749-6632.1988.tb51542.x}%
  \BibitemOpen
  \bibfield  {author} {\bibinfo {author} {\bibfnamefont {F.}~\bibnamefont
  {Bergsma}},\ }\bibfield  {title} {\enquote {\bibinfo {title} {Charm
  production measured in a 400-gev proton-copper beam dump experiment},}\
  }\href {\doibase 10.1111/j.1749-6632.1988.tb51542.x} {\bibfield  {journal}
  {\bibinfo  {journal} {Annals of the New York Academy of Sciences}\ }\textbf
  {\bibinfo {volume} {535}},\ \bibinfo {pages} {506--515} (\bibinfo {year}
  {1988})},\ \Eprint
  {http://arxiv.org/abs/https://doi.org/10.1111/j.1749-6632.1988.tb51542.x}
  {https://doi.org/10.1111/j.1749-6632.1988.tb51542.x} \BibitemShut {NoStop}%
\bibitem [{\citenamefont {{W. Martienssen and H.
  Warlimont}}(2005)}]{TantalumW2}%
  \BibitemOpen
  \bibfield  {author} {\bibinfo {author} {\bibnamefont {{W. Martienssen and H.
  Warlimont}}},\ }\enquote {\bibinfo {title} {{\textit{Springer Handbook of
  Condensed Matter and Materials Data}}},}\ \ (\bibinfo  {publisher}
  {Springer-Verlag Berlin Heidelberg},\ \bibinfo {year} {2005})\ Chap.\
  \bibinfo {chapter} {{3.1.9: Refractory metals and alloys}}, p.\ \bibinfo
  {pages} {318}\BibitemShut {NoStop}%
\bibitem [{\citenamefont {{J. M. Steichen}}(1976)}]{tungstenprops}%
  \BibitemOpen
  \bibfield  {author} {\bibinfo {author} {\bibnamefont {{J. M. Steichen}}},\
  }\bibfield  {title} {\enquote {\bibinfo {title} {{Tensile properties of
  neutron irradiated TZM and tungsten}},}\ }\href@noop {} {\bibfield  {journal}
  {\bibinfo  {journal} {J. Nucl. Mater.}\ }\textbf {\bibinfo {volume} {60}},\
  \bibinfo {pages} {13--19} (\bibinfo {year} {1976})}\BibitemShut {NoStop}%
\bibitem [{\citenamefont {\textit{et al.}}(2017)}]{Garoby_2017}%
  \BibitemOpen
  \bibfield  {author} {\bibinfo {author} {\bibfnamefont {R.~Garoby}\
  \bibnamefont {\textit{et al.}}},\ }\bibfield  {title} {\enquote {\bibinfo
  {title} {{The European Spallation Source Design}},}\ }\href {\doibase
  10.1088/1402-4896/aa9bff} {\bibfield  {journal} {\bibinfo  {journal} {Physica
  Scripta}\ }\textbf {\bibinfo {volume} {93}},\ \bibinfo {pages} {014001}
  (\bibinfo {year} {2017})}\BibitemShut {NoStop}%
\bibitem [{\citenamefont {{A. T. Nelson, J. A. O'Toole, R. A. Valicenti, S. A.
  Maloy}}(2012)}]{HIP1}%
  \BibitemOpen
  \bibfield  {author} {\bibinfo {author} {\bibnamefont {{A. T. Nelson, J. A.
  O'Toole, R. A. Valicenti, S. A. Maloy}}},\ }\bibfield  {title} {\enquote
  {\bibinfo {title} {{Fabrication of a tantalum-clad tungsten target for
  LANSCE}},}\ }\href {\doibase 10.1016/j.jnucmat.2011.11.041} {\bibfield
  {journal} {\bibinfo  {journal} {J. Nucl. Mater.}\ }\textbf {\bibinfo {volume}
  {431}},\ \bibinfo {pages} {172--184} (\bibinfo {year} {2012})}\BibitemShut
  {NoStop}%
\bibitem [{\citenamefont {{M. Kawai, K. Kikuchi, H. Kurishita, J.-F. Li, M.
  Furusaka}}(2001)}]{HIP2}%
  \BibitemOpen
  \bibfield  {author} {\bibinfo {author} {\bibnamefont {{M. Kawai, K. Kikuchi,
  H. Kurishita, J.-F. Li, M. Furusaka}}},\ }\bibfield  {title} {\enquote
  {\bibinfo {title} {{Fabrication of a tantalum-clad tungsten target for
  KENS}},}\ }\href@noop {} {\bibfield  {journal} {\bibinfo  {journal} {J. Nucl.
  Mater.}\ }\textbf {\bibinfo {volume} {296}},\ \bibinfo {pages} {312--320}
  (\bibinfo {year} {2001})}\BibitemShut {NoStop}%
\bibitem [{\citenamefont {{J. Busom Descarrega, A. T. P\'erez Fontenla, A.
  Perillo Marcone, E. Lopez Sola, M. Calviani, S. Sgobba, T. Weißg\"arber, T.
  Hutsch}}(2019)}]{HIP_Busom}%
  \BibitemOpen
  \bibfield  {author} {\bibinfo {author} {\bibnamefont {{J. Busom Descarrega,
  A. T. P\'erez Fontenla, A. Perillo Marcone, E. Lopez Sola, M. Calviani, S.
  Sgobba, T. Weißg\"arber, T. Hutsch}}},\ }\bibfield  {title} {\enquote
  {\bibinfo {title} {{Application of Hot Isostatic Pressing (HIP) technology to
  diffusion bond refractory metals for proton beam targets and absorbers at
  CERN}},}\ }\href {\doibase 10.1002/mdp2.101} {\bibfield  {journal} {\bibinfo
  {journal} {Material Design \& Processing Communications}\ } (\bibinfo {year}
  {2019}),\ 10.1002/mdp2.101}\BibitemShut {NoStop}%
\bibitem [{\citenamefont {Jones}\ and\ \citenamefont
  {Wilcox}(2018)}]{ISISclad}%
  \BibitemOpen
  \bibfield  {author} {\bibinfo {author} {\bibfnamefont {L.~G.}\ \bibnamefont
  {Jones}}\ and\ \bibinfo {author} {\bibfnamefont {D.}~\bibnamefont {Wilcox}},\
  }\bibfield  {title} {\enquote {\bibinfo {title} {{ISIS} {TS}1 project target
  {\textendash} design for manufacture},}\ }\href {\doibase
  10.1088/1742-6596/1021/1/012056} {\bibfield  {journal} {\bibinfo  {journal}
  {J. Phys.: Conference Series}\ }\textbf {\bibinfo {volume} {1021}},\ \bibinfo
  {pages} {012056} (\bibinfo {year} {2018})}\BibitemShut {NoStop}%
\bibitem [{\citenamefont {{F. Cardarelli}}(2018)}]{TantalumW}%
  \BibitemOpen
  \bibfield  {author} {\bibinfo {author} {\bibnamefont {{F. Cardarelli}}},\
  }\enquote {\bibinfo {title} {{Materials Handbook: A Concise Desktop
  Reference}},}\ \ (\bibinfo  {publisher} {Springer International Publishing},\
  \bibinfo {year} {2018})\ Chap.\ \bibinfo {chapter} {{4.3.7: Tantalum and
  tantalum alloys}}, pp.\ \bibinfo {pages} {490--513}\BibitemShut {NoStop}%
\bibitem [{\citenamefont {\textit{et al.}}(2019{\natexlab{b}})}]{sola2019beam}%
  \BibitemOpen
  \bibfield  {author} {\bibinfo {author} {\bibfnamefont {E.~Lopez~Sola}\
  \bibnamefont {\textit{et al.}}},\ }\bibfield  {title} {\enquote {\bibinfo
  {title} {{Beam impact tests of a prototype target for the Beam Dump Facility
  at CERN: experimental setup and preliminary analysis of the online
  results}},}\ }\href@noop {} {\  (\bibinfo {year} {2019}{\natexlab{b}})},\
  \Eprint {http://arxiv.org/abs/1909.07094} {arXiv:1909.07094
  [physics.ins-det]} \BibitemShut {NoStop}%
\bibitem [{\citenamefont {B{\"o}hlen}\ \emph {et~al.}(2014)\citenamefont
  {B{\"o}hlen}, \citenamefont {Cerutti}, \citenamefont {Chin}, \citenamefont
  {Fass\`o}, \citenamefont {Ferrari}, \citenamefont {Ortega}, \citenamefont
  {Mairani}, \citenamefont {Sala}, \citenamefont {Smirnov},\ and\ \citenamefont
  {Vlachoudis}}]{FLUKA_Code}%
  \BibitemOpen
  \bibfield  {author} {\bibinfo {author} {\bibfnamefont {T.T.}\ \bibnamefont
  {B{\"o}hlen}}, \bibinfo {author} {\bibfnamefont {F.}~\bibnamefont {Cerutti}},
  \bibinfo {author} {\bibfnamefont {M.P.W.}\ \bibnamefont {Chin}}, \bibinfo
  {author} {\bibfnamefont {A.}~\bibnamefont {Fass\`o}}, \bibinfo {author}
  {\bibfnamefont {A.}~\bibnamefont {Ferrari}}, \bibinfo {author} {\bibfnamefont
  {P.G.}\ \bibnamefont {Ortega}}, \bibinfo {author} {\bibfnamefont
  {A.}~\bibnamefont {Mairani}}, \bibinfo {author} {\bibfnamefont {P.R.}\
  \bibnamefont {Sala}}, \bibinfo {author} {\bibfnamefont {G.}~\bibnamefont
  {Smirnov}}, \ and\ \bibinfo {author} {\bibfnamefont {V.}~\bibnamefont
  {Vlachoudis}},\ }\bibfield  {title} {\enquote {\bibinfo {title} {The {FLUKA
  Code}: Developments and challenges for high energy and medical
  applications},}\ }\href {\doibase https://doi.org/10.1016/j.nds.2014.07.049}
  {\bibfield  {journal} {\bibinfo  {journal} {Nuclear Data Sheets}\ }\textbf
  {\bibinfo {volume} {120}},\ \bibinfo {pages} {211 -- 214} (\bibinfo {year}
  {2014})}\BibitemShut {NoStop}%
\bibitem [{\citenamefont {\textit{et al.}}(2018)}]{EDMS-ExtractionSHIP}%
  \BibitemOpen
  \bibfield  {author} {\bibinfo {author} {\bibfnamefont {Y.~Dutheil}\
  \bibnamefont {\textit{et al.}}},\ }\bibfield  {title} {\enquote {\bibinfo
  {title} {{B}eam {T}ransfer {L}ine {D}esign to the {SPS} {B}eam {D}ump
  {F}acility},}\ }in\ \href {\doibase doi:10.18429/JACoW-IPAC2018-TUPAF032}
  {\emph {\bibinfo {booktitle} {Proc. 9th International Particle Accelerator
  Conference (IPAC'18), Vancouver, BC, Canada, April 29-May 4, 2018}}},\
  \bibinfo {series and number} {\bibinfo {series} {International Particle
  Accelerator Conference}\ No.~\bibinfo {number} {9}}\ (\bibinfo  {publisher}
  {JACoW Publishing},\ \bibinfo {address} {Geneva, Switzerland},\ \bibinfo
  {year} {2018})\ pp.\ \bibinfo {pages} {751--753}\BibitemShut {NoStop}%
\bibitem [{\citenamefont {\textit{et al.}}(to be published)}]{YellowBook}%
  \BibitemOpen
  \bibfield  {author} {\bibinfo {author} {\bibfnamefont {M.~Lamont}\
  \bibnamefont {\textit{et al.}}},\ }\bibfield  {title} {\enquote {\bibinfo
  {title} {{SPS Beam Dump Facility Comprehensive Design Study}},}\ }\href@noop
  {} {\bibfield  {journal} {\bibinfo  {journal} {{CERN, Yellow Report}}\ }
  (\bibinfo {year} {to be published})}\BibitemShut {NoStop}%
\bibitem [{\citenamefont {{G. Filacchioni, E. Casagrande, U. De Angelis, G. De
  Santis, D. Ferrara}}(1994)}]{TZM_Filacchioni}%
  \BibitemOpen
  \bibfield  {author} {\bibinfo {author} {\bibnamefont {{G. Filacchioni, E.
  Casagrande, U. De Angelis, G. De Santis, D. Ferrara}}},\ }\bibfield  {title}
  {\enquote {\bibinfo {title} {{Tensile and impact properties of TZM and Mo-5\%
  Re}},}\ }\href {\doibase https://doi.org/10.1016/0022-3115(94)91037-5}
  {\bibfield  {journal} {\bibinfo  {journal} {{J. Nucl. Mater.}}\ }\textbf
  {\bibinfo {volume} {212-215, Part B}},\ \bibinfo {pages} {1288--1291}
  (\bibinfo {year} {1994})}\BibitemShut {NoStop}%
\bibitem [{\citenamefont {Schmidt}\ and\ \citenamefont
  {Ogden}(1963)}]{Tungsten_Schmidt}%
  \BibitemOpen
  \bibfield  {author} {\bibinfo {author} {\bibfnamefont {F.F.}\ \bibnamefont
  {Schmidt}}\ and\ \bibinfo {author} {\bibfnamefont {H.R.}\ \bibnamefont
  {Ogden}},\ }\bibfield  {title} {\enquote {\bibinfo {title} {The engineering
  properties of tungsten and tungsten alloys},}\ }\href@noop {} {\bibfield
  {journal} {\bibinfo  {journal} {Defense Metals Information Center, Columbus,
  Ohio}\ } (\bibinfo {year} {1963})}\BibitemShut {NoStop}%
\bibitem [{\citenamefont {{H.C. Starck}}(2017)}]{TaW_HCStarck}%
  \BibitemOpen
  \bibfield  {author} {\bibinfo {author} {\bibnamefont {{H.C. Starck}}},\
  }\bibfield  {title} {\enquote {\bibinfo {title} {{ULTRA 76 Tantalum for
  Corrosive Resistant Applications}},}\ }\href
  {https://www.hcstarck.com/ultra76_tantalum_alloy} {\bibfield  {journal}
  {\bibinfo  {journal} {Online}\ } (\bibinfo {year} {2017})}\BibitemShut
  {NoStop}%
\bibitem [{\citenamefont {{Fraunhofer IFAM}}(2017)}]{TaW_yieldFH}%
  \BibitemOpen
  \bibfield  {author} {\bibinfo {author} {\bibnamefont {{Fraunhofer IFAM}}},\
  }\bibfield  {title} {\enquote {\bibinfo {title} {{TaW-clad refractory
  metals}},}\ }\href@noop {} {\bibfield  {journal} {\bibinfo  {journal}
  {Internal report}\ } (\bibinfo {year} {2017})}\BibitemShut {NoStop}%
\bibitem [{\citenamefont {{Plansee GmbH}}(2018)}]{TZM_yieldPlansee}%
  \BibitemOpen
  \bibfield  {author} {\bibinfo {author} {\bibnamefont {{Plansee GmbH}}},\
  }\bibfield  {title} {\enquote {\bibinfo {title} {{TZM measurements}},}\
  }\href@noop {} {\bibfield  {journal} {\bibinfo  {journal} {Internal report}\
  } (\bibinfo {year} {2018})}\BibitemShut {NoStop}%
\bibitem [{\citenamefont {Zhang}\ \emph {et~al.}(2009)\citenamefont {Zhang},
  \citenamefont {Ganeev}, \citenamefont {Wang}, \citenamefont {Liu},\ and\
  \citenamefont {Alexandrov}}]{W_DBTT_Zhang}%
  \BibitemOpen
  \bibfield  {author} {\bibinfo {author} {\bibfnamefont {Y.}~\bibnamefont
  {Zhang}}, \bibinfo {author} {\bibfnamefont {A.~V.}\ \bibnamefont {Ganeev}},
  \bibinfo {author} {\bibfnamefont {J.~T.}\ \bibnamefont {Wang}}, \bibinfo
  {author} {\bibfnamefont {J.~Q.}\ \bibnamefont {Liu}}, \ and\ \bibinfo
  {author} {\bibfnamefont {I.~V.}\ \bibnamefont {Alexandrov}},\ }\bibfield
  {title} {\enquote {\bibinfo {title} {Observations on the ductile-to-brittle
  transition in ultrafine-grained tungsten of commercial purity},}\ }\href
  {\doibase https://doi.org/10.1016/j.msea.2008.07.074} {\bibfield  {journal}
  {\bibinfo  {journal} {Materials Science and Engineering: A}\ }\textbf
  {\bibinfo {volume} {503}},\ \bibinfo {pages} {37 -- 40} (\bibinfo {year}
  {2009})}\BibitemShut {NoStop}%
\bibitem [{\citenamefont {Krsjak}\ \emph {et~al.}(2014)\citenamefont {Krsjak},
  \citenamefont {Wei}, \citenamefont {Antusch},\ and\ \citenamefont
  {Dai}}]{W_DBTT_2}%
  \BibitemOpen
  \bibfield  {author} {\bibinfo {author} {\bibfnamefont {V.}~\bibnamefont
  {Krsjak}}, \bibinfo {author} {\bibfnamefont {S.H.}\ \bibnamefont {Wei}},
  \bibinfo {author} {\bibfnamefont {S.}~\bibnamefont {Antusch}}, \ and\
  \bibinfo {author} {\bibfnamefont {Y.}~\bibnamefont {Dai}},\ }\bibfield
  {title} {\enquote {\bibinfo {title} {Mechanical properties of tungsten in the
  transition temperature range},}\ }\href {\doibase
  https://doi.org/10.1016/j.jnucmat.2013.11.019} {\bibfield  {journal}
  {\bibinfo  {journal} {J. Nucl. Mater.}\ }\textbf {\bibinfo {volume} {450}},\
  \bibinfo {pages} {81 -- 87} (\bibinfo {year} {2014})}\BibitemShut {NoStop}%
\bibitem [{\citenamefont {{J. Habainy, A. Lövberg, S. Iyengar, Y. Lee, Y.
  Dai}}(2018)}]{Wfatigue}%
  \BibitemOpen
  \bibfield  {author} {\bibinfo {author} {\bibnamefont {{J. Habainy, A.
  Lövberg, S. Iyengar, Y. Lee, Y. Dai}}},\ }\bibfield  {title} {\enquote
  {\bibinfo {title} {{Fatigue properties of tungsten from two different
  processing routes}},}\ }\href {\doibase
  https://doi.org/10.1016/j.jnucmat.2017.10.061} {\bibfield  {journal}
  {\bibinfo  {journal} {{J. Nucl. Mater.}}\ }\textbf {\bibinfo {volume}
  {506}},\ \bibinfo {pages} {83--91} (\bibinfo {year} {2018})}\BibitemShut
  {NoStop}%
\bibitem [{\citenamefont {{H. Ullmaier and F. Carsughi}}(1995)}]{W_radiation}%
  \BibitemOpen
  \bibfield  {author} {\bibinfo {author} {\bibnamefont {{H. Ullmaier and F.
  Carsughi}}},\ }\bibfield  {title} {\enquote {\bibinfo {title} {{Radiation
  damage problems in high power spallation neutron sources}},}\ }\href
  {\doibase https://doi.org/10.1016/0168-583X(95)00590-0} {\bibfield  {journal}
  {\bibinfo  {journal} {{Nucl. Instrum. Methods Phys. Res. B}}\ }\textbf
  {\bibinfo {volume} {101}},\ \bibinfo {pages} {406--421} (\bibinfo {year}
  {1995})}\BibitemShut {NoStop}%
\bibitem [{\citenamefont {M{\"u}ller}\ \emph {et~al.}(1994)\citenamefont
  {M{\"u}ller}, \citenamefont {Gavillet}, \citenamefont {Victoria},\ and\
  \citenamefont {Martin}}]{TZM_radiation}%
  \BibitemOpen
  \bibfield  {author} {\bibinfo {author} {\bibfnamefont {G.V.}\ \bibnamefont
  {M{\"u}ller}}, \bibinfo {author} {\bibfnamefont {D.}~\bibnamefont
  {Gavillet}}, \bibinfo {author} {\bibfnamefont {M.}~\bibnamefont {Victoria}},
  \ and\ \bibinfo {author} {\bibfnamefont {J.L.}\ \bibnamefont {Martin}},\
  }\bibfield  {title} {\enquote {\bibinfo {title} {{Postirradiation tensile
  properties of Mo and Mo alloys irradiated with 600 MeV protons}},}\
  }\href@noop {} {\bibfield  {journal} {\bibinfo  {journal} {J. Nucl. Mater.}\
  }\textbf {\bibinfo {volume} {212}},\ \bibinfo {pages} {1283--1287} (\bibinfo
  {year} {1994})}\BibitemShut {NoStop}%
\bibitem [{\citenamefont {Byun}\ and\ \citenamefont
  {Maloy}(2008)}]{Ta_radiation_BYUN}%
  \BibitemOpen
  \bibfield  {author} {\bibinfo {author} {\bibfnamefont {T.~S.}\ \bibnamefont
  {Byun}}\ and\ \bibinfo {author} {\bibfnamefont {S.~A.}\ \bibnamefont
  {Maloy}},\ }\bibfield  {title} {\enquote {\bibinfo {title} {Dose dependence
  of mechanical properties in tantalum and tantalum alloys after low
  temperature irradiation},}\ }\href@noop {} {\bibfield  {journal} {\bibinfo
  {journal} {J. Nucl. Mater.}\ }\textbf {\bibinfo {volume} {377}},\ \bibinfo
  {pages} {72--79} (\bibinfo {year} {2008})}\BibitemShut {NoStop}%
\bibitem [{\citenamefont {Chen}\ \emph {et~al.}(2001)\citenamefont {Chen},
  \citenamefont {Ullmaier}, \citenamefont {Flo{\ss}dorf}, \citenamefont
  {K{\"u}hnlein}, \citenamefont {Duwe}, \citenamefont {Carsughi},\ and\
  \citenamefont {Broome}}]{Ta_radiation_CHEN}%
  \BibitemOpen
  \bibfield  {author} {\bibinfo {author} {\bibfnamefont {J.}~\bibnamefont
  {Chen}}, \bibinfo {author} {\bibfnamefont {H.}~\bibnamefont {Ullmaier}},
  \bibinfo {author} {\bibfnamefont {T.}~\bibnamefont {Flo{\ss}dorf}}, \bibinfo
  {author} {\bibfnamefont {W.}~\bibnamefont {K{\"u}hnlein}}, \bibinfo {author}
  {\bibfnamefont {R.}~\bibnamefont {Duwe}}, \bibinfo {author} {\bibfnamefont
  {F.}~\bibnamefont {Carsughi}}, \ and\ \bibinfo {author} {\bibfnamefont
  {T.}~\bibnamefont {Broome}},\ }\bibfield  {title} {\enquote {\bibinfo {title}
  {{Mechanical properties of pure tantalum after 800 MeV proton
  irradiation}},}\ }\href@noop {} {\bibfield  {journal} {\bibinfo  {journal}
  {J. Nucl. Mater.}\ }\textbf {\bibinfo {volume} {298}},\ \bibinfo {pages}
  {248--254} (\bibinfo {year} {2001})}\BibitemShut {NoStop}%
\bibitem [{\citenamefont {J.~Habainy}(2018)}]{TC_tungsten}%
  \BibitemOpen
  \bibfield  {author} {\bibinfo {author} {\bibfnamefont {Y.~Lee Y.~Dai}\
  \bibnamefont {J.~Habainy}, \bibfnamefont {S.~Iyengar}},\ }\bibfield  {title}
  {\enquote {\bibinfo {title} {Thermal diffusivity of tungsten irradiated with
  protons up to 5.8 dpa},}\ }\href {\doibase
  https://doi.org/10.1016/j.jnucmat.2018.06.041} {\bibfield  {journal}
  {\bibinfo  {journal} {J. Nucl. Mater.}\ }\textbf {\bibinfo {volume} {509}},\
  \bibinfo {pages} {152--157} (\bibinfo {year} {2018})}\BibitemShut {NoStop}%
\bibitem [{\citenamefont {Hurh}(2017)}]{Hurh:IPAC2017-WEOCB3}%
  \BibitemOpen
  \bibfield  {author} {\bibinfo {author} {\bibfnamefont {P.}~\bibnamefont
  {Hurh}},\ }\bibfield  {title} {\enquote {\bibinfo {title} {{The Radiation
  Damage in Accelerator Target Environments (RaDIATE) Collaboration R\&D
  Program {-} Status and Future Activities}},}\ }in\ \href {\doibase
  https://doi.org/10.18429/JACoW-IPAC2017-WEOCB3} {\emph {\bibinfo {booktitle}
  {Proc. of International Particle Accelerator Conference (IPAC'17),
  Copenhagen, Denmark, 14-19 May, 2017}}},\ \bibinfo {series and number}
  {\bibinfo {series} {International Particle Accelerator Conference}\
  No.~\bibinfo {number} {8}}\ (\bibinfo  {publisher} {JACoW},\ \bibinfo
  {address} {Geneva, Switzerland},\ \bibinfo {year} {2017})\ pp.\ \bibinfo
  {pages} {2550--2553},\ \bibinfo {note}
  {https://doi.org/10.18429/JACoW-IPAC2017-WEOCB3}\BibitemShut {NoStop}%
\bibitem [{\citenamefont {Ammigan}\ \emph {et~al.}(2017)\citenamefont {Ammigan}
  \emph {et~al.}}]{Ammigan:IPAC2017-WEPVA138}%
  \BibitemOpen
  \bibfield  {author} {\bibinfo {author} {\bibfnamefont {K.}~\bibnamefont
  {Ammigan}} \emph {et~al.},\ }\bibfield  {title} {\enquote {\bibinfo {title}
  {{The RaDIATE High-Energy Proton Materials Irradiation Experiment at the
  Brookhaven Linac Isotope Producer Facility}},}\ }in\ \href {\doibase
  https://doi.org/10.18429/JACoW-IPAC2017-WEPVA138} {\emph {\bibinfo
  {booktitle} {Proc. of International Particle Accelerator Conference
  (IPAC'17), Copenhagen, Denmark, 14-19 May, 2017}}},\ \bibinfo {series and
  number} {\bibinfo {series} {International Particle Accelerator Conference}\
  No.~\bibinfo {number} {8}}\ (\bibinfo  {publisher} {JACoW},\ \bibinfo
  {address} {Geneva, Switzerland},\ \bibinfo {year} {2017})\ pp.\ \bibinfo
  {pages} {3593--3596},\ \bibinfo {note}
  {https://doi.org/10.18429/JACoW-IPAC2017-WEPVA138}\BibitemShut {NoStop}%
\bibitem [{\citenamefont {{R.M. Christensen}}(2007)}]{Christensen}%
  \BibitemOpen
  \bibfield  {author} {\bibinfo {author} {\bibnamefont {{R.M. Christensen}}},\
  }\bibfield  {title} {\enquote {\bibinfo {title} {{A Comprehensive Theory of
  Yielding and Failure for Isotropic Materials}},}\ }\href {\doibase
  10.1115/1.2712847} {\bibfield  {journal} {\bibinfo  {journal} {Journal of
  Engineering Materials and Technology}\ }\textbf {\bibinfo {volume} {129}},\
  \bibinfo {pages} {173--181} (\bibinfo {year} {2007})}\BibitemShut {NoStop}%
\bibitem [{\citenamefont {{H. A. Calderon, G.Kostorz,
  G.Ullrich}}(1993)}]{TZMfatigue}%
  \BibitemOpen
  \bibfield  {author} {\bibinfo {author} {\bibnamefont {{H. A. Calderon,
  G.Kostorz, G.Ullrich}}},\ }\bibfield  {title} {\enquote {\bibinfo {title}
  {{Microstructure and plasticity of two molybdenum-base alloys (TZM)}},}\
  }\href {\doibase https://doi.org/10.1016/0921-5093(93)90447-M} {\bibfield
  {journal} {\bibinfo  {journal} {{Materials Science and Engineering: A}}\
  }\textbf {\bibinfo {volume} {160}},\ \bibinfo {pages} {189 -- 199} (\bibinfo
  {year} {1993})}\BibitemShut {NoStop}%
\bibitem [{\citenamefont {{R. I. Stephens, A. Fatemi, et al.}}(2000)}]{Fatemi}%
  \BibitemOpen
  \bibfield  {author} {\bibinfo {author} {\bibnamefont {{R. I. Stephens, A.
  Fatemi, et al.}}},\ }\enquote {\bibinfo {title} {{\textit{Metal Fatigue in
  Engineering, 2nd Edition}}},}\ \ (\bibinfo  {publisher}
  {Wiley-Interscience},\ \bibinfo {year} {2000})\ Chap.\ \bibinfo {chapter}
  {{10.3: Stress-based criteria}}, p.\ \bibinfo {pages} {325}\BibitemShut
  {NoStop}%
\bibitem [{\citenamefont {{J. Habainy, S. Iyengar, K. B. Surreddi, Y. Lee, Y.
  Dai}}(2018)}]{WO3_oxi}%
  \BibitemOpen
  \bibfield  {author} {\bibinfo {author} {\bibnamefont {{J. Habainy, S.
  Iyengar, K. B. Surreddi, Y. Lee, Y. Dai}}},\ }\bibfield  {title} {\enquote
  {\bibinfo {title} {Formation of oxide layers on tungsten at low oxygen
  partial pressures},}\ }\href {https://doi.org/10.1016/j.jnucmat.2017.12.018}
  {\bibfield  {journal} {\bibinfo  {journal} {J. Nucl. Mater.}\ }\textbf
  {\bibinfo {volume} {506}},\ \bibinfo {pages} {26--34} (\bibinfo {year}
  {2018})}\BibitemShut {NoStop}%
\bibitem [{\citenamefont {{D. J. S. Findlay, G. P. {\v{S}}koro, G. M. Allen, D.
  J. Haynes, D. M. Jenkins, D. Wilcox}}(2018)}]{Decay_RAL}%
  \BibitemOpen
  \bibfield  {author} {\bibinfo {author} {\bibnamefont {{D. J. S. Findlay, G.
  P. {\v{S}}koro, G. M. Allen, D. J. Haynes, D. M. Jenkins, D. Wilcox}}},\
  }\bibfield  {title} {\enquote {\bibinfo {title} {Measurement and calculation
  of decay heat in {ISIS} spallation neutron target},}\ }\href
  {https://doi.org/10.1016/j.nima.2018.08.007} {\bibfield  {journal} {\bibinfo
  {journal} {Nucl. Instrum. Methods Phys. Res Sec. A}\ }\textbf {\bibinfo
  {volume} {908}},\ \bibinfo {pages} {91--96} (\bibinfo {year}
  {2018})}\BibitemShut {NoStop}%
\bibitem [{\citenamefont {{Plansee GmbH}}(2017)}]{Production_plansee}%
  \BibitemOpen
  \bibfield  {author} {\bibinfo {author} {\bibnamefont {{Plansee GmbH}}},\
  }\bibfield  {title} {\enquote {\bibinfo {title} {{Manufacturing of refractory
  metals}},}\ }\href@noop {} {\bibfield  {journal} {\bibinfo  {journal}
  {Internal communication}\ } (\bibinfo {year} {2017})}\BibitemShut {NoStop}%
\end{thebibliography}%

\end{document}